\documentclass{tlpmod}

\usepackage[english]{babel}
\usepackage{latexsym}
\usepackage{epsfig}
\usepackage{url}

\newcommand{\IMPL}[0]{\,\supset\,}
\newcommand{\IFF}[0]{\,\equiv\,}
\def\AND     { \,\wedge\,		  }
\def\OR      { \,\vee\,			  }

\newtheorem{definition}{Definition}
\newtheorem{proposition}{Proposition}
\newtheorem{theorem}{Theorem}
\newtheorem{lemma}{Lemma}
\newtheorem{corollary}{Corollary}
\newtheorem{example}{Example}

\newcommand{\tuple}[1]{\langle #1 \rangle}
\newcommand{\HT}[0]{\mathsf{HT}}
\renewcommand{\HT}[0]{\mathrm{HT}}

\newcommand{\tsim}[0]{\hspace{1.1mm}\rule{.16mm}{2.3mm}\hspace{-.3mm}
\raisebox{.1ex}{$\sim$}\hspace{.8mm}}
\renewcommand{\tsim}[0]{\rule{.11mm}{2.3mm}\hspace{-.3mm}
\raisebox{.1ex}{$\sim$}}
\newcommand{\nmseq}[2]{\;\tsim_{#1}^{\mbox{\scriptsize\sc #2}}\,}

\renewcommand{\nmseq}[1]{\;\tsim_{#1}\,}
\newcommand{\brave}[0]{\nmseq{\mbox{\scriptsize\sc brave}}}
\newcommand{\skept}[0]{\nmseq{\mbox{\scriptsize\sc skept}}}
\renewcommand{\brave}[0]{\nmseq{b}}
\renewcommand{\skept}[0]{\nmseq{s}}

\newcommand{\Trans}[1]{{\cal S}[#1]}
\newcommand{\Transv}[1]{{\cal V}[#1]}

\newcommand{\clneg}[1]{\overline{#1}}

\newcommand{\LPif}{\leftarrow}

\newcommand{\quantifier}{{\sf Q}}
\newcommand{\MIN}[0]{\ensuremath{\mathit{min}}}
\newcommand{\MAX}[0]{\ensuremath{\mathit{max}}}
\newcommand{\F}{\ensuremath{{\cal F}}}
\renewcommand{\F}{\ensuremath{{\cal I}}}

\newcommand{\Iht}{\ensuremath{\langle I_H, I_T \rangle}}
\newcommand{\I}[0]{\ensuremath{\nu}}
\newcommand{\Ivo}[0]{\I}
\newcommand{\val}[2]{\Ivo_{#1}({#2})}
\newcommand{\valF}[2]{\val{\F}{#1,#2}}
\newcommand{\valFs}[2]{\val{\overline{\F}}{#1,#2}}
\newcommand{\commadots}[0]{,\ldots ,}
\newcommand{\iec}[0]{i.e.,\ }

\newcommand{\egc}[0]{e.g.,\ }

\newcommand{\eqmod}[2]{\ensuremath{\mathcal{M}[#1,#2]}}

\newcommand{\lc}[1]{\ensuremath{\mathit{lc}(#1)}}
\newcommand{\lb}[0]{\ensuremath{\mathrm{lb}}}
\newcommand{\lcn}[1]{\ensuremath{\mathit{lc}^+(#1)}}
\newcommand{\var}[1]{\mathit{var}(#1)}
\newcommand{\PhiT}[1]{\ensuremath{\tau[#1]}}
\newcommand{\PhiTpp}[1]{\ensuremath{\tau''[#1]}}
\newcommand{\PhiTn}[1]{\ensuremath{\tau^\star[#1]}}
\newcommand{\PhiHT}[1]{\ensuremath{{\cal T}_{\mathit{HT}}[#1]}}
\newcommand{\PhiE}[1]{\ensuremath{{\cal T}_E[#1]}}

\newcommand{\PhiEn}[1]{\ensuremath{{\cal T}_{\mathit{S}}[#1]}}

\newcommand{\sub}[2]{[#1 / #2]}
\newcommand{\Circ}[1]{\ensuremath{\mathrm{CIRC}(#1)}}
\newcommand{\ICirc}[1]{\ensuremath{\mathrm{ICIRC}(#1)}}
\newcommand{\subn}[4]{[#1 / #2 \commadots #3 / #4]}

\newcommand{\TLP}[1]{{\cal T}_{\mathit{DLP}}[#1]}

\newcommand{\semprop}{{\tt sem\-prop}}
\newcommand{\qpro}{{\tt qpro}}
\newcommand{\qube}{{\tt qube-bj}}
\newcommand{\quantor}{{\tt quantor}}

\newcommand{\Pol}{{\rm P}}
\newcommand{\NP}{{\rm NP}}

\newcommand{\CONP}{\mbox{\rm co-}\NP }
\renewcommand{\CONP}{\mbox{\rm co-}\NP }
\newcommand{\co}{\mbox{\rm co-}}
\newcommand{\SigmaP}[1]{{\Sigma}_{#1}^{P}}
\newcommand{\PiP}[1]{{\Pi}_{#1}^{P}}

\newcommand{\DeltaP}[1]{{\Delta}_{#1}^{P}}

\newcommand{\negf}{{\mathord{\sim}}}

\newcommand{\PSPACE}{{\rm PSPACE}}

\title[Characterising Equilibrium Logic and Nested Logic Programs]{Characterising Equilibrium Logic and Nested Logic Programs: Reductions and Complexity\thanks{%
Part of the material in this paper  appeared in preliminary form
in the Proceedings of the 10th Portuguese Conference on Artificial Intelligence (EPIA 2001), pp.\ 306-320, LNCS vol.~2258, Springer, 2001.}\thanks{The second author dedicates this paper to the memory of his father, who died in October 2008.}}
\author[D.\ Pearce, H.\ Tompits, and S.\ Woltran]
{DAVID PEARCE\\
Universidad Polit\'{e}cnica de Madrid,\\
Dep.\ de Inteligencia Artificial, Grupo CLIP\\
E-28660 Boadilla del Monte {\rm (}Madrid{\rm )}, Spain   \\
\email{david.pearce@upm.es}
\and HANS TOMPITS
\\
           Technische Universit\"{a}t
           Wien,\\
Institut f{\"u}r Informationssysteme 184/3,\\
Arbeitsbereich Wissensbasierte Systeme, \\
Favoritenstrasse 9-11,  A-1040 Vienna, Austria\\
           \email{tompits@kr.tuwien.ac.at}
\and STEFAN WOLTRAN
\\
           Technische Universit\"{a}t
           Wien,\\
Institut f{\"u}r Informationssysteme 184/2, \\
Arbeitsbereich Datenbanken und Artificial Intelligence, \\
Favoritenstrasse 9-11,  A-1040 Vienna, Austria\\
           \email{woltran@dbai.tuwien.ac.at}}

\submitted{11 January 2007}
 \revised{15 April 2009}
 \accepted{5 May 2009}

\begin{document}

\maketitle

\begin{abstract}
Equilibrium logic is an approach to nonmonotonic reasoning that 
extends the stable-model and 
answer-set semantics for logic programs. 
In particular, it includes the general case of \emph{nested logic programs}, where arbitrary Boolean combinations are permitted in heads and bodies of rules, as special kinds of theories.
In this paper, we present polynomial reductions of the main reasoning tasks associated with equilibrium logic and nested logic programs into \emph{quantified propositional logic}, an extension of classical propositional logic where quantifications over atomic formulas are permitted.
Thus, quantified propositional logic is a fragment of second-order logic, and its formulas are usually referred to as \emph{quantified Boolean formulas} (QBFs).
We provide reductions not only for decision problems, but also for the central semantical concepts of equilibrium logic and nested logic programs.
In particular, our encodings map a given decision problem into some QBF such that the latter is valid precisely in case the former holds.
The basic tasks we deal with here are the \emph{consistency problem}, \emph{brave reasoning}, and \emph{skeptical reasoning}.
Additionally, we also provide encodings for testing equivalence of theories or programs under different notions of equivalence, viz.\ \emph{ordinary}, \emph{strong}, and \emph{uniform equivalence}.
For all considered reasoning tasks, we analyse their computational complexity and give strict complexity bounds.
Hereby, our encodings yield upper bounds in a direct manner.
Besides this useful feature, our approach has the following benefits:
First, our encodings yield a \emph{uniform axiomatisation} for 
a variety of problems in a common language.
Secondly, extant solvers for QBFs can be used as back-end inference engines to realise implementations of the encoded task in a rapid prototyping manner.
Thirdly, our axiomatisations also allow us to straightforwardly relate equilibrium logic with circumscription.
\end{abstract}

\begin{keywords}
answer-set programming, equilibrium logic, reduction techniques, quantified Boolean formulas, computational complexity, equivalence testing
\end{keywords}

\section{Introduction}\label{sec:intro}

\emph{Equilibrium logic}, introduced by \citeN{Pearce97},
 is a general purpose 
propositional formalism for nonmonotonic reasoning.
It is a form of minimal-model reasoning in the non-classical logic of \emph{here-and-there}, which is basically intuitionistic logic restricted to two worlds, ``here'' and ``there''.
One of the main features 
of equilibrium logic is that, under all the usual classes of logic programs, 
it is equivalent to reasoning under answer-set semantics and therefore amounts 
to a conservative extension of answer-set inference to the full propositional 
language. 
It even includes the general case
of \emph{nested logic programs} \cite{Lifschitz99},
where arbitrary Boolean expressions are permitted in heads and bodies
of program rules. 
With the emergence of answer-set solvers such as 
\texttt{DLV} \cite{leon-etal-2002-dlv}, 
\texttt{Smodels} \cite{simo-etal-2002}, or
\texttt{ASSat} \cite{Lin02a}, 
\emph{answer-set programming} (ASP) now provides a practical and 
viable environment for knowledge representation and declarative problem 
solving.
(For an overview of equilibrium logic, see \citeN{Pearce06}; an excellent treatise on ASP is the comprehensive textbook  by 
\citeN{bara-2002}; a survey of applications is compiled by \citeN{wasp-wp5}.)

Our main contribution in this work is to 
provide a {uniform axiomatisation} of 
various decision problems associated with equilibrium logic
and nested logic programs.
In fact, we  
reformulate 
these problems 
within a common language such that a
sentence in that language is valid iff it encodes a yes-instance of the 
given decision problem. 
Moreover, the resulting sentences are not only useful for decision 
problems, but also characterise via models 
certain semantical concepts in the original setting of 
equilibrium logic and stable semantics.
Finally, our axiomatisations allow us to draw in a direct manner upper complexity bounds for the decision problems in question.
We strengthen these results by providing matching lower bounds, thus giving \emph{strict} complexity results for the encoded problems.
At the same time, these complexity results show that our axiomatisations are \emph{adequate} in the
sense of~\citeN{Besnard05}, \iec roughly speaking, they reflect the inherent complexity of the original decision problems.

The target language for our endeavour is 
\emph{quantified propositional logic}, 
an extension of classical propositional logic
where quantifications over atomic formulas are permitted.
Thus,
quantified propositional logic is
a fragment of second-order logic, and its sentences are usually referred to as
\emph{quantified Boolean formulas} (QBFs).
We describe polynomial-time constructible encodings providing
characterisations of the problems under consideration 
by means of QBFs.
Moreover, since logic programming under the answer-set semantics
is just a special form of equilibrium logic, our encodings 
for the latter yield also encodings for the former. 
Although the sizes of the general encodings are already super-linear in the sizes of the input
problems, and linear in case of programs as inputs, we provide optimised encodings for logic programs by eliminating some redundancies.

The general approach of using reductions to QBFs as adopted here has already been applied to several diverse reasoning problems from the area of artificial intelligence.
Examples include the following fields: 
\begin{itemize}
\item nonmonotonic formalisms like logic-based abduction, default logic, and modal nonmonotonic logics \cite{Egly00c,Eiter02,Tompits03};
\item consistency-based belief revision \cite{Delgrande04}; \item paraconsistent reasoning \cite{ariden03,Besnard05}; and 
\item planning \cite{Rintanen99b}.
\end{itemize}

Concerning the specific encodings discussed in this paper, for both equilibrium logic and logic programs with nested expressions
we provide encodings for the basic reasoning tasks associated with these formalisms.
To wit, we consider the \emph{consistency problem}, \emph{brave reasoning}, and \emph{skeptical reasoning}.
In the context of equilibrium logic, these tasks are the following:
\begin{itemize}
\item decide whether a given theory possesses some equilibrium model;
\item decide whether some equilibrium model of a given theory satisfies a given formula; and 
\item decide whether all equilibrium models of a given theory satisfy a given formula.
\end{itemize}
The corresponding tasks for logic programs are defined analogously.
Additionally, we also provide encodings for testing the equivalence of theories or programs under different notions of equivalence.
Besides ordinary equivalence, which, in case of equilibrium logic, tests whether two theories have the same equilibrium models, we also consider more refined ones, viz.\ \emph{strong equivalence} \cite{Lifschitz01} and \emph{uniform equivalence} \cite{Eiter03,Pearce04}.
In detail, two theories are 
\begin{itemize}
\item strongly equivalent iff, for 
any addition of formulas, the two augmented theories
are ordinarily equivalent, and
\item uniformly equivalent iff, for 
any addition of \emph{atoms}, the two augmented theories
are ordinarily equivalent. 
\end{itemize}
(Again, the above concepts are defined with respect to theories of equilibrium logic; for logic programs, the corresponding notions are defined \emph{mutatis mutandis}.)
Intuitively, while strong equivalence realises a form of substitution principle, uniform equivalence, first studied in the context of datalog programs~\cite{Maher88,Sagiv88}, is useful for dealing with hierarchically structured program components.
Importantly, these notions serve as formal underpinnings for program simplification and modular programming \cite{Eiter03a,Pearce04b}.

As equilibrium logic is based on the logic of here-and-there, we also
address the latter in our work, not only because
it underlies equilibrium logic, but also since it is closely related to strong and uniform equivalence.
Indeed, a module representing reasoning in 
here-and-there plays a central role in all axiomatisations we provide.
This module, together with other simple modules, will serve as sort of a  
tool box for arranging the encodings of the different problems considered.
Hence, our reductions to quantified propositional logic provide an elegant axiomatisation of the considered reasoning tasks in a uniform setting.

Regarding the concrete complexity results discussed in this paper, we show that the consistency problem and brave reasoning are $\SigmaP{2}$-complete, while skeptical reasoning is $\PiP{2}$-complete. Furthermore, checking ordinary or uniform equivalence is $\PiP{2}$-complete, while checking strong equivalence is $\CONP$-complete. 
All these results hold for both equilibrium logic as well as for nested logic programs.
This shows that, with respect to the considered reasoning tasks, equilibrium logic and nested logic programs behave complexitywise precisely as disjunctive logic programs.

Let us now have some words about the advantages of our reduction approach to quantified propositional logic.
First of all, as already pointed out above, reductions to QBFs allow us to axiomatise diverse problems in a uniform language. As well, they enable us to derive upper complexity bounds with ease.
Secondly, an increasing number of practicably efficient solvers for quantified propositional logic emerged in recent years, like, \egc \semprop~\cite{Letz02}, 
\qube~\cite{Giunchiglia03}, \quantor~\cite{Biere04}, and \qpro~\cite{Egly06}.\footnote{See \url{www.qbflib.org} for an up-to-date listing of available QBF solvers.}
These can be used as back-end inference engines for our encoded tasks, allowing us to build implementations for the considered reasoning tasks in a rapid-prototyping manner.
The viability of using QBF solvers for computing different knowledge-representation tasks was already demonstrated in initial work along these lines (see, \egc \citeN{Egly00c}; a more recent experimental evaluation is discussed, \egc by Oetsch et al.\ \citeyear{Oetsch06c,Oetsch07c}).
Thirdly, our axiomatisations allow us also to relate equilibrium logic to circumscription~\cite{McCarthy80}.
For the case of disjunctive logic programs, an early 
result on the relation between answer sets and 
circumscription was given by \citeN{Lin91}.
Our encodings generalise \citeANP{Lin91}'s result 
not only to 
logic programs with nested expressions 
but also to full equilibrium logic.
We note that similar generalisations are independently provided by \citeN{Ferraris:Lee:Lifschitz:07}.

The remainder of the paper is laid out as follows. 
First, we review 
the nonclassical logic of here-and-there and its nonmonotonic extension, 
equilibrium logic. 
Then, we introduce logic programs as a special case
of this logic
and discuss the basic elements of quantified propositional logic.
We then turn in Section~3 to the main issue of the paper, 
the characterisation of equilibrium logic in terms of QBFs.
As an underlying task, we show how to re-express satisfiability in 
here-and-there in the setting of classical logic. 
We then address logic programs with nested expressions, providing reductions which are optimisations of the ones for equilibrium logic.
We continue with relating our approach to circumscription, and the final part of Section~3 deals with the encodings for
ordinary, uniform, and strong equivalence. 
In Section~4, we turn to the complexity of the considered reasoning tasks, 
and Section~5 briefly discusses issues which concern the addition of a 
second negation, which is usually termed \emph{strong negation} or (as a misnomer) \emph{classical
negation}. Section~6 addresses related work and Section~7 concludes the paper with a brief discussion.
A more involving proof is relegated to an appendix.

\section{Background}\label{sec:logical}

Throughout this paper, we use a propositional language whose alphabet 
consists of (i)~the {primitive logical connectives}
$\neg$, $\vee$, $\wedge$, and 
$\IMPL$, (ii)~the {logical constants} $\top$ and $\bot$, (iii)~the {punctuation symbols} `$($' and `$)$', and (iv)~a class of ({propositional}) {variables}.
The class of ({propositional}) {formulas} is constructed in the usual inductive fashion.
Propositional variables are also referred to as \emph{atomic formulas}, or simply \emph{atoms}.
We use $p,q,r,\ldots$ to denote propositional variables and Greek lower-case letters to denote arbitrary formulas; in either case, the letters may also contain subscripts.
In order to improve the readability of formulas, we also allow the use of other kinds of punctuation symbols besides `$($' and `$)$'. 
A propositional formula 
which does not contain the sentential connective $\IMPL$ 
is called an \emph{expression}.
As usual, a \emph{literal} 
is either a propositional variable (a \emph{positive literal}) or a propositional variable preceded by the negation symbol $\neg$ (a \emph{negative literal}).
Furthermore, $\varphi\IFF\psi$ is an abbreviation of the 
formula $(\varphi\IMPL\psi)\AND(\psi\IMPL\varphi)$.

The \emph{logical complexity}, $\lc{\varphi}$, of a formula $\varphi$ is 
the number of occurrences of the logical symbols $\neg$, $\vee$, $\wedge$, and
$\IMPL$ in~$\varphi$.

By $\var{\varphi}$ we understand the set of all variables occurring in a formula $\varphi$.
Likewise, for a set $T$ of formulas, $\var{T}$ refers to the set of all variables occurring in the formulas in $T$, \iec $\var{T}=\bigcup_{\varphi\in T}\var{\varphi}$.

A finite set of formulas is called a \emph{theory}.
Usually, a theory $T$ will be identified with a conjunction $\bigwedge_{\varphi\in T}\varphi$ 
of its elements.
For $T=\emptyset$, we define $\bigwedge_{\varphi \in T}\varphi = \top$.

By a (\emph{classical}) \emph{interpretation}, $I$, we understand a set of variables.
Informally, a variable $p$ is true under $I$ iff $p\in I$.
The \emph{truth value} of a formula $\varphi$, in the sense of classical propositional logic, is denoted by $\val{I}{\varphi}$ and is defined in the usual way.
We write $\models\varphi$ to indicate that $\varphi$ is valid in classical propositional logic, 
and, accordingly, we write $T\models \phi$ to denote
that $\phi$ is a logical consequence of $T$.

We employ the following notational convention: 
Given a formula $\varphi$, by $\varphi'$ we understand the result of replacing each variable $p$ occurring in $\varphi$ by a globally new variable $p'$.
This is applied analogously to sets of formulas.
In particular, if $I$ is an interpretation, then $I'$ is the interpretation $\{p'\mid p\in I\}$ which is disjoint to $I$.
Whenever needed, we apply this concept iteratively, \iec 
we also use new variables $p''$, $p'''$, etc.

\subsection{Equilibrium Logic}\label{sec:eq-logic-back}

We recall the basic concepts of equilibrium logic, an approach to nonmonotonic reasoning developed by \citeN{Pearce97} as  a generalisation of the answer-set semantics for logic programs.
We give only the relevant aspects here; for more details, the reader is referred to Pearce~\shortcite{Pearce97,Pearce99,Pearce06}, Pearce et al.~\shortcite{Pearce00,Pearce00a}, Lifschitz et al.~\shortcite{Lifschitz01}, 
and the logic texts cited below.
Besides the propositional version discussed here, a first-order variant of equilibrium logic has been introduced as well~\cite{pearce-valverde-sl05}.

Equilibrium logic is based on the non-classical logic of \emph{here-and-there}, which we denote by $\HT$. 
The language of $\HT$ is given by the class of propositional formulas as described above, 
and the axioms and rules of inference of $\HT$ 
are those of intuitionistic logic (cf., \egc \citeN{Da86}) together with  the axiom schema 
$$
(\neg \varphi \IMPL  \psi ) \IMPL
(((\psi \IMPL \varphi ) \IMPL \psi ) \IMPL \psi)
$$
which characterises the three-valued here-and-there logic of \citeN{Heyting30} and \citeN{Goedel32} (hence, $\HT$ is sometimes known as \emph{G\"odel's three-valued logic}).
The standard version of equilibrium logic has two kinds of negation, \emph{intuitionistic negation}, $\neg$, and \emph{strong negation}, $\negf$.
For simplicity, we deal first with the restricted version containing just the first negation.
Later, in Section~\ref{sec:strong}, we show how strong negation can be added.

The model theory of $\HT$ is based on the usual Kripke semantics for intuitionistic logic, which is given in terms of Kripke frames of form $\langle W, \leq \rangle$,
where $W$ is a set of \emph{points}, or \emph{worlds}, and $\leq$ is a partial-ordering on $W$ (cf., \egc \citeN{Da86}),
except that Kripke frames for $\HT$ are restricted to those containing exactly two worlds, say $H$ (``here'') and $T$ (``there''), with $H \leq T$.
As in ordinary Kripke semantics for intuitionistic logic, we can imagine that in each world a set of atoms is verified and that, once verified ``here'', an atom remains verified ``there''. 

In view of the restricted nature of Kripke frames for $\HT$, it is convenient to define the semantics of $\HT$ in terms of \emph{HT-interpretations}, which are ordered pairs of form $\Iht$, where $I_H$ and $I_T$ are sets of variables such that 
$I_H \subseteq I_T$.
For an HT-interpretation $\F=\Iht$, a world $w\in\{H,T\}$, and a formula $\varphi$, the \emph{truth value}, $\valF{w}{\varphi} \in \{ 0 , 1 \}$, \emph{of $\varphi$ in $w$ under $\F$} is given as follows:

\begin{enumerate}
\item if $\varphi=\top$, then $\valF{w}{\varphi} = 1$;

\item if $\varphi=\bot$, then $\valF{w}{\varphi} = 0$;

\item if $\varphi=p$, for some variable $p$, then $\valF{w}{\varphi} = 1$ if
      $p\in I_w$, and $\valF{w}{\varphi} = 0$ otherwise;

\item if $\varphi=\neg \psi$, then
      $\valF{w}{\varphi} = 1$ if, for every world $u$ such that $w\leq u$, 
	$\valF{u}{\psi} = 0$, 
      and $\valF{w}{\varphi} = 0$ otherwise;

\item if $\varphi=(\varphi_1\AND \varphi_2)$, then
      $\valF{w}{\varphi} = \MIN(\{\valF{w}{\varphi_1},\valF{w}{\varphi_2}\})$;

\item if $\varphi=(\varphi_1\OR \varphi_2)$, then
      $\valF{w}{\varphi} = \MAX(\{\valF{w}{\varphi_1},\valF{w}{\varphi_2}\})$; and

\item if $\varphi=(\varphi_1\IMPL \varphi_2)$, then
      $\valF{w}{\varphi} = 1$ if, for every world $u$ such that $w\leq u$, 
	$\valF{u}{\varphi_1}\leq\valF{u}{\varphi_2}$,  
      and $\valF{w}{\varphi} = 0$ otherwise.
\end{enumerate}

We say that $\varphi$ is \emph{true under $\F$ in $w$} if $\valF{w}{\varphi}=1$,
otherwise $\varphi$ is \emph{false under $\F$ in $w$}. An HT-interpretation $\F=\Iht$ \emph{satisfies} $\varphi$, or $\F$ is an \emph{HT-model} of $\varphi$, iff $\valF{H}{\varphi}=1$. 
If $\varphi$ possesses some HT-interpretation satisfying it, then $\varphi$ is said to be \emph{HT-satisfiable}, and if every HT-interpretation satisfies $\varphi$, then $\varphi$ is \emph{HT-valid}. 
An HT-interpretation is an HT-model of a set $T$ of formulas iff it is an HT-model of all elements of $T$.
Finally, an HT-interpretation $\Iht$ is said to be \emph{total} if $I_H = I_T$, and \emph{non-total} 
otherwise (\iec if $I_H\subset I_T$).

It is easily seen that any HT-valid formula is valid in classical logic, but the converse does not always hold. For instance, $p\OR\neg p$ and $\neg\neg p\IMPL p$ are valid in classical logic but not in the logic of here-and-there, because $\F=\langle \emptyset,\{p\}\rangle$ is not an HT-model for either of these formulas. 

We say that
two theories are \emph{equivalent in the logic
of here-and-there}, or \emph{HT-equivalent}, iff they
possess the same HT-models. Two formulas, $\varphi$ and $\psi$, are HT-equivalent iff
the theories $\{\varphi\}$ and $\{\psi\}$ are HT-equivalent.

Equilibrium logic is characterised in terms of a particular minimal-model construction in HT.  Formally, an \emph{equilibrium model} of a theory $T$ is a total HT-interpretation $\langle I,I \rangle$ such that (i)~$\langle I,I \rangle$ is an HT-model of $T$, and (ii)~for every proper subset $J$ of $I$, $\langle J,I \rangle$ is not an HT-model of $T$.
$\tuple{I,I}$ is an equilibrium model of a formula $\varphi$ iff $\tuple{I,I}$ is an equilibrium model of $\{\varphi\}$.
 
A formula $\varphi$ is a \emph{brave consequence} of a theory $T$, symbolically $T\brave\varphi$, iff some equilibrium model of $T$ satisfies $\varphi$.
Dually, $\varphi$ is a \emph{skeptical consequence} of $T$, symbolically $T\skept\varphi$, iff all equilibrium models of $T$ satisfy $\varphi$.

The basic reasoning tasks in the context of equilibrium logic are the following decision problems:

\begin{itemize}
\item Decide whether a given theory $T$ possesses some equilibrium model.

\item Given a theory $T$ and a formula $\varphi$, decide whether $T\brave\varphi$ holds.

\item Given a theory $T$ and a formula $\varphi$, decide whether $T\skept\varphi$ holds.

\end{itemize}

The first task is called the \emph{consistency problem}; the second and third tasks are respectively called \emph{brave reasoning} and \emph{skeptical reasoning}. 

The following two propositions are straightforward and will be useful later on:

\begin{proposition} \label{lemma:HT}
For any HT-interpretation $\F=\langle I_H, I_T\rangle$ and any propositional 
formula $\varphi$, the following relations hold:
\begin{enumerate}
\item\label{lemma:HT:1}
 $\valF{T}{\varphi}=1$ iff $\val{I_T}{\varphi}=1$;
\item\label{lemma:HT:2} $\valF{H}{\varphi}=1$ implies $\valF{T}{\varphi}=1$; and
\item\label{lemma:HT:3}
$\valF{H}{\varphi}=1$ iff $\val{I_H}{\varphi}=1$, if  $\varphi$ is an expression (\iec a formula without $\IMPL$) that does not contain  negation.
\end{enumerate}
\end{proposition}

Notice that the first part of this proposition states that $\varphi$ is true under $\F=\langle I_H, I_T\rangle$ in the world $T$ iff ${\varphi}$ is true under $I_T$ in classical logic. The second part is a direct consequence of 
the notion of an HT-interpretation, viz.\ in view of the proviso $I_H\subseteq I_T$, which holds for each
HT-interpretation $\Iht$.
The third part states that formulas without negations and implications
can be evaluated by pure classical means, \iec in both worlds separately.

\begin{proposition} \label{lemma:total}
A total HT-interpretation $\langle I,I \rangle$ is an HT-model of $\varphi$ iff $I$ is a model of $\varphi$ in classical logic.
\end{proposition}

\subsection{Logic Programs}\label{sec:lp-background}

Next, we review logic programs with nested expressions under the stable-model semantics, following \citeN{Lifschitz99}.\footnote{Analogously as for equilibrium logic, here we consider programs with only one kind of negation first, however, 
corresponding to default negation; the case of strong negation will be discussed later.} 
Programs of this kind are characterised by being comprised of
rules whose bodies and heads are expressions, \iec formulas composed of $\AND$, $\OR$, and $\neg$ only.

Formally, by a \emph{rule}, $r$, we understand an ordered pair of the form 
\[
H(r) \LPif B(r),
\]
where $B(r)$ and $H(r)$ are expressions.
We call $B(r)$ the \emph{body} of $r$ and $H(r)$ the \emph{head} of $r$. A \emph{nested logic program}, or simply a \emph{program}, $\Pi$, is a finite set of rules.
A \emph{fact} is a rule of form $p\LPif$ where $p$ is an atom.

We employ for rules and programs the same notational convention concerning priming  as we did for formulas, \iec $r'$ shall be the result of replacing each atom $p$ in $r$ by $p'$, and, similarly, $\Pi'$ is given by $\{r'\mid r\in\Pi\}$.

Note that programs properly generalise \emph{normal logic programs}, which are characterised by the condition that bodies of rules are conjunctions of literals 
and heads are just atoms, and \emph{disjunctive logic programs},
which are similarly defined except that heads may be disjunctions of atoms.

In what follows, we associate to each rule $r$ a corresponding propositional 
formula 
\[
\hat{r}=B(r)\IMPL H(r)
\]
and, accordingly, to each program $\Pi$ a corresponding set of formulas 
\[
\hat{\Pi}=\{\hat{r}\mid r\in\Pi\}.
\]
Furthermore, we define $\var{\Pi}$ as the set of all variables occurring in $\Pi$, \iec $\var{\Pi}=\var{\hat{\Pi}}$.

We call expressions, rules, and programs \emph{basic} iff they do not contain the operator~$\neg$. An interpretation $I$ is a \emph{model} of a basic program $\Pi$ if it is a model of the associated set $\hat{\Pi}$ of formulas.

Given an interpretation $I$ and an (arbitrary) program $\Pi$, the \emph{reduct}, $\Pi^I$, of $\Pi$ with respect to $I$ is the basic program obtained from $\Pi$ by replacing every occurrence of an expression $\neg \psi$ in $\Pi$ which is not in the scope of any other negation by $\top$ if $\neg\psi$ is true under $I$ (\iec if $\val{I}{\neg\psi}=1$), and by $\bot$ otherwise. $I$ is a \emph{stable model} of $\Pi$ iff it is a minimal model (with respect to set inclusion) of the reduct $\Pi^I$.

\begin{example}\label{ex:lp}
Consider a logic program $\Pi$ consisting of the single rule
\begin{equation}\label{eq:ex:lp}
p \LPif (q\AND r)\OR(\neg q \AND \neg s).
\end{equation}
Let us check whether $I=\{p\}$ is a stable model of $\Pi$.
Since both $\neg q$ and $\neg s$ are true under $I$, we obtain
$$
\Pi^I=\{p\LPif 
(q\AND r)\OR(\top\AND\top)\}.
$$
Since $(q\AND r)\OR(\top\AND\top)$ is classically equivalent to $\top$, 
the only minimal model of $\Pi^I$ is $\{p\}$.
So, $I$ is a stable model of $\Pi$. In fact, there is no other
stable model of $\Pi$.
\end{example}

A formula $\varphi$ 
is said to be a \emph{brave consequence} of a logic
program $\Pi$, symbolically $\Pi\brave\varphi$,  iff there is a stable model $I$ of $\Pi$ such that $\varphi$ is true under $I$, and 
$\varphi$ is a \emph{skeptical consequence} of $\Pi$, symbolically $\Pi\skept\varphi$, iff $\varphi$ is
true under all stable models $I$ of $\Pi$. 

The basic reasoning tasks in the context of logic programs are defined \emph{mutatis mutandis} as for equilibrium logic and are likewise referred to as the \emph{consistency problem}, \emph{brave reasoning}, and \emph{skeptical reasoning}, respectively.

It is well known that disjunctive logic programs satisfy the so-called \emph{anti-chain property}, which expresses that, for all stable models $I$ and $J$ of a program $\Pi$, if $I\subseteq J$, then $I=J$.
The next example shows that programs with nested expressions do not meet the anti-chain property in general.

\begin{example}\label{ex:4}
Let $\Pi = \{ p \LPif \neg\neg p\}$. We show that
$\Pi$ possesses two stable models, viz.\
$I_1=\emptyset$ and
$I_2=\{p\}$. 
Concerning $I_1$, we have that
$\neg p$ is true under $I_1$, and therefore
$\Pi^{I_1}=\{p \LPif \bot\}$, which has $\emptyset$ as
its minimal model.
As for $I_2$, here it holds that
$\neg p$ is false under $I_2$, and so
$\Pi^{I_2}=\{p \LPif \top\}$, which has $\{p\}$ as its minimal model.
Consequently, we have stable models $I_1$ and $I_2$ of $\Pi$ such that
$I_1\subseteq I_2$ but $I_1\neq I_2$.
Hence, the anti-chain property is violated.
\end{example}

The following result, originally established by \citeN{Pearce97} and later generalised to nested programs by \citeN{Lifschitz01}, reveals the close connection between equilibrium models and stable models, showing that stable models are actually a special case of equilibrium models.

\begin{proposition}\label{prop:eqmsm}
For any program $\Pi$, $I$ is a stable model of $\Pi$ iff $\langle I, I \rangle$ is an equilibrium model of $\hat{\Pi}$.
\end{proposition}

\subsection{Quantified Propositional Logic}

We now introduce 
\emph{quantified propositional logic}, an extension of classical
propositional logic in which formulas are permitted to contain quantifications over propositional variables.
More formally, in addition to the symbols used to construct propositional formulas, the language of quantified propositional logic 
is assumed to contain
 the two symbols $\forall$ and $\exists$.
The formation rules for constructing formulas of quantified propositional logic 
are similar to the usual formation rules for propositional formulas, together with the condition that if $\Phi$ is a formula and $p$ is a propositional variable, then $(\forall p\,\Phi)$ and $(\exists p\,\Phi)$ are also formulas.
We call $\forall p$ a \emph{universal quantifier} and $\exists p$ an \emph{existential quantifier}, for every variable $p$.
Informally, $(\forall p\,\Phi)$ expresses that $\Phi$ is true for \emph{all} truth assignments of $p$, while $(\exists p\,\Phi)$ means that $\Phi$ is true for \emph{some} truth assignment of $p$.
We allow the usual convention of omitting parentheses if no ambiguity arises.
Formulas of quantified propositional logic are usually referred to as  \emph{quantified Boolean formulas} (QBFs) and are denoted by Greek upper-case letters (recall that we use Greek lower-case letters to denote standard propositional formulas).
We define the \emph{logical complexity} of a QBF $\Phi$
analogously to the case of ordinary propositional formulas,
\iec $\lc{\Phi}$ is given as the number of occurrences of the logical symbols $\forall$, $\exists$,
$\neg$, $\vee$, $\wedge$, and $\IMPL$ in~$\Phi$.

The semantics of quantified propositional logic is defined as follows.
First, we require some ancillary notation.
An occurrence of a variable $p$ in a QBF $\Phi$ is \emph{bound} 
if it is either within the scope of a quantifier 
$\quantifier p$ ($\quantifier\in\{\forall,\exists\}$) or else is immediately preceded by the symbol $\forall$ or $\exists$.
An occurrence of $p$ in $\Phi$ which is not bound is called \emph{free}. 
We say that $p$ has a \emph{free occurrence} in $\Phi$, or that $p$ is a \emph{free variable} of $\Phi$, if at least one occurrence of $p$ in $\Phi$ is free.
If $\Phi$ contains no free variables, then $\Phi$ is 
\emph{closed}, otherwise $\Phi$ is \emph{open}. 
Furthermore, 
$\Phi \subn{p_1}{\psi_1}{p_n}{\psi_n}$ denotes the 
result of uniformly substituting the
free occurrences of variables $p_i$ in $\Phi$ by $\psi_i$ ($1\leq i\leq n$).
This notation is extended to sets of variables and formulas in the obvious way.
That is to say, for pairwise disjoint sets $V_i=\{p_1^i\commadots p_{j_i}^i\}$ 
of variables 
and sets $S_i=\{\psi_1^i\commadots \psi_{j_i}^i\}$ of formulas, for $1\leq i \leq n$, $\Phi \subn{V_1}{S_1}{V_n}{S_n}$ stands for
\[
\Phi[p_1^1/\psi_1^1\commadots p_{j_1}^1/\psi_{j_1}^1\commadots p_1^n/\psi_1^n\commadots p_{j_n}^n/\psi_{j_n}^n].
\]

Given a (classical) interpretation $I$ and a QBF $\Phi$, the \emph{truth value $\val{I}{\Phi}$ of $\Phi$ under $I$} 
is inductively defined as follows:
\begin{enumerate}
\item if $\Phi=\top$, then $\val{I}{\Phi} = 1$;

\item if $\Phi=\bot$, then $\val{I}{\Phi} = 0$;

\item if $\Phi=p$, for some variable $p$, then $\val{I}{\Phi} = 1$ if 
      $p\in I$, and $\val{I}{\Phi} = 0$ otherwise;

\item if $\Phi=\neg \Psi$, then 
      $\val{I}{\Phi} = 1 - \val{I}{\Psi}$;

\item if $\Phi=(\Phi_1\AND \Phi_2)$, then 
      $\val{I}{\Phi} = \MIN(\{\val{I}{\Phi_1},\val{I}{\Phi_2}\})$;

\item if $\Phi=(\Phi_1\OR \Phi_2)$, then 
      $\val{I}{\Phi} = \MAX(\{\val{I}{\Phi_1},\val{I}{\Phi_2}\})$;

\item if $\Phi=(\Phi_1\IMPL \Phi_2)$, then 
      $\val{I}{\Phi} = 1$ iff $\val{I}{\Phi_1} \leq \val{I}{\Phi_2}$;

\item if $\Phi= \forall p \, \Psi$, then 
      $\val{I}{\Phi} =  \MIN(\val{I}{\Psi \sub{p}{\top}},\val{I}{\Psi \sub{p}{\bot}})$; and

\item if $\Phi= \exists  p \, \Psi$, then 
      $\val{I}{\Phi} =  \MAX(\val{I}{\Psi \sub{p}{\top}},\val{I}{\Psi \sub{p}{\bot}})$.
 \end{enumerate}
Observe that it obviously holds that
\[
\val{I}{\forall p \, \Psi} =  \val{I}{\Psi \sub{p}{\top}\AND\Psi \sub{p}{\bot}}
 \quad \mbox{and} \quad
\val{I}{\exists  p \, \Psi} =  \val{I}{\Psi \sub{p}{\top}\OR\Psi \sub{p}{\bot}}.
\]

We say that $\Phi$ is \emph{true under $I$} if $\val{I}{\Phi}=1$, 
otherwise $\Phi$ is \emph{false under $I$}. 
If $\val{I}{\Phi}=1$, then $I$ is a \emph{model} of $\Phi$. 
Likewise, for a set $S$ of QBFs, if $\val{I}{\Phi}=1$ for all $\Phi\in S$,
then $I$ is a model of $S$.
If $\Phi$ has some model, then $\Phi$ is said to be \emph{satisfiable}.
If $\Phi$ is true under any interpretation, then $\Phi$ is \emph{valid}, in symbols $\models\Phi$.
Observe that a closed QBF is either valid or unsatisfiable, because closed QBFs are either true under each interpretation or false under each interpretation. 
Hence, for closed QBFs, there is no need to refer to particular interpretations.

In the sequel, we employ the following abbreviations in the context of QBFs: 
Let $S=\{\varphi_1,\ldots,\varphi_n\}$ and $T=\{\psi_1,\ldots,\psi_n\}$ be 
sets of indexed formulas. 
Then, 
$S\leq T$ abbreviates 
$\bigwedge_{i=1}^n (\varphi_i \IMPL \psi_i)$, and $S < T$ is a 
shorthand for 
$(S\leq T) \AND \neg (T\leq S)$.
Furthermore, for a set $P = \{ p_1,\ldots,p_n\}$ of propositional 
variables and a quantifier $\quantifier\in\{\forall,\exists\}$, we 
let $\quantifier P \, \Phi$ stand for the formula 
$\quantifier p_1 \quantifier p_2\cdots\quantifier p_n\,\Phi$.

The operators $\leq$ and $<$ are fundamental tools for expressing certain tests on sets of atoms.
In particular, the following properties hold:
Let $P=\{p_1\commadots p_n\}$ be a set of indexed atoms, and let $I_1\subseteq P$ and $I_2\subseteq P$ be two interpretations.
Then, 
\begin{enumerate}
\item[(i)] $I_1\cup I_2'$ is a model of $P\leq P'$ iff $I_1\subseteq I_2$, and 
\item[(ii)] $I_1\cup I_2'$ is a model of $P< P'$ iff $I_1\subset I_2$.
\end{enumerate}

We also note the following obvious but central property, which will be relevant later on: 
Let $\Phi$ be a QBF whose free variables are given by $P$, and let 
$Q\subseteq P$. 
Then, an interpretation $I\subseteq P\setminus Q$ is a model of 
$\exists Q\Phi$ iff there is some 
$J\subseteq Q$ such that $J\cup I$ is a model of $\Phi$.

Similar to classical first-order logic, there are several results concerning 
the shifting and renaming of quantifiers.
We list some fundamental relations below and refer the interested reader to
\citeN{Egly03a} and \citeN{Woltran03} for a fuller discussion.

\begin{proposition}\label{prop:quantifier-shift}
Let $p$ and $q$ be atoms, and $\quantifier\in\{\forall,\exists\}$.
Furthermore, let
$\Phi$ and $\Psi$ be QBFs such that $\Psi$ does not contain free occurrences
of $p$.
Then, 
\begin{enumerate}
\item
$\models (\neg \exists p\, \Phi) \IFF \forall p  (\neg \Phi)$,
\item
$\models (\neg \forall p\,\Phi) \IFF \exists p (\neg \Phi)$,
\item
$\models (\Psi \circ \quantifier p\,\Phi) \IFF \quantifier 
p(\Psi \circ \Phi)$, for $\circ\in\{\AND,\OR,\IMPL\}$,
and 
\item
$\models (\quantifier q\, \Psi)\IFF(\quantifier p\,\Psi\sub{q}{p})$.
\end{enumerate}
\end{proposition}
A QBF $\Phi$ is in \emph{prenex normal form} iff
it is of the form
$$
\quantifier_1 V_1  \quantifier_2 V_2 \ldots \quantifier_n V_n\, \phi,
$$
where
$\phi$ is a  propositional formula,
$\quantifier_i \in \{\exists,\forall\}$ such that $\quantifier_i\neq\quantifier_{i+1}$  
for $1\leq i \leq n-1$, and
$V_1\commadots V_n$ are pairwise disjoint sets
of propositional variables.
Without going into details, we mention that any QBF is easily transformed
into an equivalent QBF in prenex normal form (by applying, among other reduction 
steps, the equivalences depicted in Proposition~\ref{prop:quantifier-shift}).
In fact, this transformation can be carried out in polynomial time.

Historically, among the first logical analyses of systems dealing with quantifiers over propositional variables are the investigations by \citeN{Russel06} (``theory of implication'') and by \citeN{Luk30}
(``erweiterter Aussagenkalk{\"u}l''), not to mention the monumental \emph{Principia Mathematica}~\cite{PM}.
The particular idea of quantifying propositional variables was extended by \citeN{Lesniewski29} in his \emph{protothetic logic} where  variables  whose values are {\em truth functions} are allowed and
quantification is defined over these variables (cf.\ also \citeN{Srzednicki98} for more details about \citeANP{Lesniewski29}'s system).\footnote{%
A more elaborate overview on these early historical aspects of
propositional quantification can be found
in~\S28
of \citeN{Church56}.}
At the beginning of the 1970s, propositional quantification
came into the spotlight of computer science---in particular in the
emerging field of complexity theory~\cite{Garey79}---when evaluation problems for QBFs were recognised as the
prototypical problems for the
\emph{polynomial hierarchy}~\cite{Stockmeyer76} as well 
as for the prominent complexity class $\PSPACE$~\cite{Meyer73}.
Details on this issue  
are given 
Section~\ref{sec:complexity}.

\section{Characterisations}\label{sec:charact}

We now show how equilibrium logic and nested logic programs can be mapped in polynomial time into QBFs.
We first deal with the case of equilibrium logic, and afterwards we provide optimised encodings for logic programs.
Then, we shed some light on the relation between our encodings and circumscription~\cite{McCarthy80}, thereby discussing an extension of a well-known result by \citeN{Lin91}.
We conclude this section by expressing different notions of equivalence in terms of ordinary and quantified propositional logic, respectively.
Note that, throughout this paper, we deal with theories and programs which are defined as \emph{finite} sets, in view of the finiteness of QBFs.

\subsection{Encodings for Equilibrium Logic}\label{sec:eqlogic-char}

Our first goal is to express satisfiability in the logic of here-and-there in terms of satisfiability in classical logic. We begin with the following translation:

\begin{definition}\label{def:encoding}
Let $\varphi$ be a formula. Then, $\PhiT{\varphi}$ is inductively defined as follows:
\begin{enumerate}
\item if $\varphi$ is an atomic formula, or one of $\top$ or $\bot$, then $\PhiT{\varphi} = \varphi$,
\item if $\varphi=(\varphi_1 \circ \varphi_2)$, for $\circ \in \{\AND,\OR\}$, then $\PhiT{\varphi}=\PhiT{\varphi_1}\circ\PhiT{\varphi_2}$,
\item if $\varphi=\neg\psi$, then $\PhiT{\varphi} = 
\neg \psi'$, and
\item if $\varphi=(\varphi_1 \IMPL \varphi_2)$, then $\PhiT{\varphi} = (\PhiT{\varphi_1} \IMPL \PhiT{\varphi_2}) \AND (\varphi_1' \IMPL \varphi_2')$.
\end{enumerate}
\end{definition}

Recall that, as per our convention, primed formulas refer to the result of replacing each atom $p$ occurring in the corresponding unprimed formula by a globally new atom $p'$. 

Intuitively, this translation encodes the inductive truth conditions of HT, where
the primed formulas in $\PhiT{\varphi}$ correspond to formulas evaluated in the world ``there'' and the unprimed formulas correspond to formulas evaluated in ``here''. 
While the connectives $\AND$ and $\OR$ behave in each world like in classical logic, and hence $\PhiT{\varphi}$ just accesses inductively the translations of the direct subformulas of $\varphi$ in these cases, the treatment of the remaining connectives $\neg$ and $\IMPL$ involves \emph{different worlds}, which is achieved in 
$\PhiT{\varphi}$ by making use of primed formulas.
As satisfaction of a formula in HT is determined at world $H$, and the latter accesses both itself and the world $T$, the inductive evaluation must proceed in $H$, whereas the world $T$ is accessible to no other world than itself, so evaluating a formula in ``there'', the inductive process comes to a halt.
This mechanism is directly evident for the case where $\varphi=(\varphi_1 \IMPL \varphi_2)$:
Here, the part $(\PhiT{\varphi_1} \IMPL \PhiT{\varphi_2})$ represents the inductive evaluation of the truth conditions in $H$ whereas the part $(\varphi_1' \IMPL \varphi_2')$ evaluates $\varphi$ in world $T$.
In principle, this also holds for the case where $\varphi=\neg\psi$, but the definition of $\PhiT{\varphi}$ involves actually an optimisation of the general pattern:
To wit, in view of Part~\ref{lemma:HT:2} of Proposition~\ref{lemma:HT}, the truth value of $\neg\psi$ is completely determined at world ``there'': if $\psi$ is false in $T$, it must also be false in $H$,
so $\neg\psi$ is true in $H$, and if $\psi$ is true in $T$, it directly follows that $\neg\psi$ is false in $H$.
Hence, no inductive clause is necessary.
We remark that this optimisation for $\neg$ was not realised in the preliminary version of this paper~\cite{Pearce01}, where $\PhiT{\neg\psi}$ was defined as $\neg \PhiT{\psi} \AND 
\neg \psi'$.

We proceed by estimating the size of $\PhiT{\varphi}$ in terms of the size of the input formula~$\varphi$. 

\begin{proposition}\label{lemma:length}
Let $\PhiT{\cdot}$ be the transformation defined above.
Then,
\begin{enumerate}
\item $\lc{\PhiT{\varphi}} \leq \lc{\varphi} + (\lc{\varphi}+1)\cdot
 \lb(\lc{\varphi}+1)$, for any formula $\varphi$, and

\item\label{lemma:length:2}
$\lc{\PhiT{\varphi}} = \lc{\varphi}$, for any expression $\varphi$.
\end{enumerate}
\end{proposition}

\begin{proof}
Part 1 is shown
by a tedious yet straightforward induction on $\lc{\varphi}$, and 
Part 2 follows by definition.
\end{proof}

Hence, $\lc{\PhiT{\varphi}}$ is super-linear in $\lc{\varphi}$ and linear in $\lc{\varphi}$ if $\varphi$ is an expression.

\begin{example}\label{example-upper}
To illustrate the upper bound in Part~1, consider the formula
$$
\varphi = (p_1\IMPL p_2) \IMPL (p_3 \IMPL p_4)
$$
for which we have that
\[
\begin{array}{l}
\PhiT{\varphi}  = 
\Big(
\big(
(p_1\IMPL p_2) \AND (p'_1\IMPL p'_2)\big)
\IMPL \big(
(p_3 \IMPL p_4) \AND (p'_3 \IMPL p'_4)
\big)
\Big) 
\AND
\\[1ex] 
\qquad
\qquad\big((p'_1\IMPL p'_2) \IMPL (p'_3 \IMPL p'_4)\big).
\end{array}
\]
Now, $\lc{\varphi}=3$, $\lc{\PhiT{\varphi}}=11$, and
$$
\lc{\varphi} + (\lc{\varphi}+1)\cdot \lb(\lc{\varphi}+1)=
3+4\cdot \lb 4=3+4\cdot 2=
11,
$$
so $\lc{\PhiT{\varphi}}$ matches the upper bound.
\end{example}

In order to fully express the semantics of the logic of here-and-there, in addition to encoding its inductive truth conditions, we must also make sure that all formulas true  ``here'' are also true  ``there''.
However, this can be conveniently expressed in terms of the condition $V\leq V'$, where $V$ is the set of atoms occurring in $\varphi$. Indeed, we have the following relation:

\begin{lemma} \label{thm:HTSat}
Let $\varphi$ be a formula and $V=\var{\varphi}$.
Furthermore, let $I_H,I_T\subseteq V$ be interpretations. 
Then, $\langle I_H, I_T\rangle$ is an HT-model of $\varphi$ iff 
$I_H \cup I'_T$ is a model of $$\PhiHT{\varphi}=(V\leq V')\AND\PhiT{\varphi}.$$ 
\end{lemma}

\begin{proof} 
See \ref{app:proof:HT}. 
\end{proof}

Note that, for $V$ as in the above lemma, 
$$\lc{V\leq V'}=2|\var{\varphi}|-1,$$ 
and since $|\var{\varphi}|\leq\lc{\varphi}+1$ always holds, 
we obtain that 
\[
\begin{array}{r@{~}c@{~}l}
\lc{\PhiHT{\varphi}} &=& \lc{V\leq V'}+\lc{\PhiT{\varphi}}+1\\[1ex]
&\leq& 2\lc{\varphi}+\lc{\PhiT{\varphi}}+2=2(\lc{\varphi}+1)+\lc{\PhiT{\varphi}}.
\end{array}
\]
Therefore, in virtue of Proposition~\ref{lemma:length}, 
it follows that
\[
\begin{array}{r@{~}c@{~}l}
\lc{\PhiHT{\varphi}}
&\leq& 2(\lc{\varphi}+1)+\lc{\varphi} + (\lc{\varphi}+1)
\lb(\lc{\varphi}+1)\\[1ex]
&=&3(\lc{\varphi}+1)+ (\lc{\varphi}+1)\lb(\lc{\varphi}+1)-1.
\end{array}
\]
That is,
$\lc{\PhiHT{\varphi}}$ is of the same order as $\lc{\PhiT{\varphi}}$, viz.\ super-linear.
Note that for the formula $\varphi$ from Example~\ref{example-upper} the upper bound in the above inequality is again reached.

\begin{example}\label{ex:1}
To illustrate the mechanism of transformation $\PhiHT{\cdot}$, consider the formula $\varphi = \neg\neg p \IMPL p$. We already demonstrated in Section~\ref{sec:eq-logic-back} that $\varphi$ is not
valid in the logic of here-and-there, although it is clearly a tautology of 
classical propositional logic. 

Let us first construct $\PhiT{\varphi}$, which is given by
$$
(\PhiT{\neg \neg p} \IMPL \PhiT{p}) \AND (\neg \neg p' \IMPL p').
$$
Evaluating 
$\PhiT{\neg\neg p}$ and $\PhiT{p}$, we get
\begin{equation}\label{eq:ex1}
( \neg \neg p' \IMPL p) \AND (\neg \neg p' \IMPL p').
\end{equation}
The first conjunct of (\ref{eq:ex1}) is equivalent to
$
(p' \IMPL p)
$ in classical propositional logic
and the second conjunct is a tautology of classical propositional logic.
Hence,  the entire translation $\PhiHT{\varphi}=(p \IMPL p') \AND \PhiT{\varphi}$ is equivalent to
$$
(p \IMPL p') \AND (p' \IMPL p),
$$
which
has two models, viz.\  $I_1=\emptyset$ and $I_2=\{p,p'\}$.
Therefore, by Lemma~\ref{thm:HTSat},
the HT-models of $(\neg\neg p \IMPL p)$ are given by $\langle \emptyset,\emptyset\rangle$ and $\langle\{p\},\{p\}\rangle$.
Observe that $\langle \emptyset,\{p\}\rangle$ is therefore not an HT-model of $\varphi$, which is in accordance with our discussion above.
\end{example}

We proceed with equilibrium models. 
Since these kinds of models are HT-models subject to an additional minimality condition, we can extend our encoding for HT taking this aspect into account.
Now, quantified propositional logic gives us the means to deal with minimality conditions in an elegant manner, by employing existential quantifiers in combination with negation and the comparison operator $<$.
In particular, the formula 
$$\Phi=\varphi'\AND\neg\exists V\big((V < V') \AND \varphi \big)$$
expresses that, for any model $I'$ of $\Phi$, $I$ is a minimal model of $\varphi$, \iec $I$ is a model of $\varphi$ and there is no $J\subset I$ such that $J$ is also a model of $\varphi$.\footnote{Actually, the formula $\Phi$ is closely related to circumscription~\cite{McCarthy80}, which we examine in more detail in Section~\ref{sec:circ}.}
Just factoring in that we have to deal with HT-interpretations, we thus arrive at the following encoding:

\begin{theorem} \label{thm:eqm}
Let $\varphi$ be a formula, $V=\var{\varphi}$, and $I\subseteq V$.
Then, $\langle I,I\rangle$ is an equilibrium model of $\varphi$ iff $I'$ is a model of
$$
\PhiE{\varphi}= \varphi' \AND \neg \exists V \big( (V < V') \AND \PhiT{\varphi} \big).
$$
\end{theorem}

\begin{proof}
We first note that $\PhiE{\varphi}$ is obviously equivalent to 
\begin{equation}\label{eq:E}
\varphi' \AND \neg \exists V \big( (V < V') \AND \PhiHT{\varphi} \big),
\end{equation}
because $(V < V')$ is logically equivalent to
$(V < V') \AND (V \leq V')$.

Let $I\subseteq V$ be some interpretation. Recall that $\langle I,I\rangle$ is an equilibrium model of $\varphi$ iff (i)~$\langle I,I\rangle$ is an HT-model of $\varphi$, and (ii) for every
$J\subset I$, $\langle J,I \rangle$ is  not an HT-model of $\varphi$.
We show that (i) and (ii) hold iff $I'$ is a model of (\ref{eq:E}).

First of all, by Proposition~\ref{lemma:total} and a simple renaming, it follows that Condition (i) holds iff $\varphi'$ is true under $I'$. Now consider 
the QBF $\Psi=\exists V ( (V < V') \AND \PhiHT{\varphi} )$. By the properties of $<$ and the semantics of the existential quantifier, we have that $\Psi$ is true under $I'$ iff there is some $J\subset I$ such that $\PhiHT{\varphi}$ is true under $J\cup I'$. 
Hence, invoking Lemma~\ref{thm:HTSat}, we get that $\Psi$ is true under $I'$ iff Condition~(ii) does not hold. Consequently, (ii) holds iff 
$\neg \Psi$ is true under $I'$. Therefore, (i) and (ii) are jointly satisfied iff $I'$ is a model of 
$\varphi' \AND \neg \Psi$.
\end{proof}

Observe that, as for ${\PhiHT{\cdot}}$, the logical complexity of ${\PhiE{\cdot}}$ is again of super-linear order.

\begin{example}\label{ex:2}
Let us compute the equilibrium models of $\neg\neg p \IMPL p$ by means of Theorem~\ref{thm:eqm}.
From Example~\ref{ex:1}, we already know that 
$\PhiT{\varphi}$ is equivalent to $p' \IMPL p$. 
Moreover, $V<V'$ stands in the present case for the formula
$(p \IMPL p') \AND \neg (p' \IMPL p)$.
Hence, $\PhiE{\varphi}$ is equivalent to
\begin{equation}\label{eq3}
(\neg\neg p' \IMPL p') \AND \neg \exists p \Big( (p \IMPL p') \AND \neg (p' \IMPL p) \AND (p' \IMPL p)\Big).
\end{equation}
Note that $\neg (p' \IMPL p) \AND (p' \IMPL p)$ makes the whole formula in 
the scope of $\exists p$ unsatisfiable, so 
(\ref{eq3}) is equivalent to 
$(\neg \neg p' \IMPL p')$,
which is a tautology of classical logic. 
Thus, every subset of $V'=\{p'\}$ is a model of $\PhiE{\varphi}$, and Theorem~\ref{thm:eqm} implies that the equilibrium models of $(\neg \neg p \IMPL p)$ are given by $\langle \emptyset,\emptyset\rangle$ and 
$\langle \{p\},\{p\} \rangle$.  
\end{example}

Having now the encoding $\PhiE{\cdot}$ at hand, we easily obtain corresponding encodings for the basic reasoning tasks associated with equilibrium logic.
Recall that we identify theories with a conjunction of their (finitely many) elements.

\begin{corollary} \label{cor:eqcons}
Let $T$ be a theory, $\varphi$ a formula, $V=\var{T}$, and $W=\var{T\cup\{\varphi\}}$.
Then, 

\begin{enumerate}
\item $T$ has an equilibrium model iff $\models \exists V'\, \PhiE{T}$,

\item $T\brave\varphi$ iff $\models\exists W'\, \big( \PhiE{T} \AND \varphi'\big)$, and

\item $T\skept\varphi$ iff $\models\forall W'\, \big( \PhiE{T} \IMPL \varphi'\big)$.

\end{enumerate}

\end{corollary}

\subsection{Encodings for Logic Programs}

In view of Proposition~\ref{prop:eqmsm},
our encodings introduced so far, based on the transformation $\PhiE{\cdot}$,  yield corresponding encodings for logic programs in a straightforward way.
For instance, by Corollary~\ref{cor:eqcons}, we have that $\Pi$ has a stable model iff $\exists V'\, \PhiE{\hat{\Pi}}$ is valid, for $V=\var{\Pi}$.
These encodings, then, have logical complexities which are \emph{linear} in the size of the input programs, in view of the fact that only expressions occur in programs (cf.\ Part~\ref{lemma:length:2} of Proposition~\ref{lemma:length}).
However, it turns out that
we can still eliminate some redundancy in $\PhiE{\cdot}$ for the case of 
logic programs, which is inherited from the general handling of implication in $\PhiE{\cdot}$, yielding a more compact transformation than $\PhiE{\hat{\Pi}}$.

\begin{theorem} \label{thm:stable:models}
Let $\Pi$ be a logic program, 
$\hat{\Pi}=\{B(r)\IMPL H(r)\mid r\in\Pi\}$ the set of 
formulas associated with $\Pi$, $V=\var{\Pi}$, and $I\subseteq V$.
Then, $I$ is a stable model of $\Pi$ iff $I'$ is a model of 
$$
\PhiEn{\Pi} =  
\hat{\Pi}'
\AND 
\neg \exists V \Big(
(V < V') \AND 
\bigwedge_{r \in \Pi} \!
\big(
\PhiT{B(r)} \IMPL \PhiT{H(r)} \big)
\Big).
$$
\end{theorem}

\begin{proof}
We show that $\PhiEn{\Pi}$ is equivalent to
$$
\PhiE{\hat{\Pi}}= \hat{\Pi}' \AND \neg \exists V \big( (V < V') \AND \PhiT{\hat{\Pi}} \big).
$$
Then, the result immediately holds by Theorem~\ref{thm:eqm} and Proposition~\ref{prop:eqmsm}. 

First of all, by definition of $\PhiT{\cdot}$, 
\[
\PhiT{\hat{\Pi}}=\bigwedge_{r\in\Pi}\Big(\big(\PhiT{B(r)}\IMPL\PhiT{H(r)}\big)\AND
\big(B(r')\IMPL H(r')\big)\Big).
\]
By associativity of $\wedge$, and by identifying 
$\bigwedge_{r\in\Pi}\big(B(r')\IMPL H(r')\big)$ with $\hat{\Pi}'$, we have that
$\PhiT{\hat{\Pi}}$ can be written as
\[
\bigwedge_{r\in\Pi}\big(\PhiT{B(r)}\IMPL\PhiT{H(r)}\big)\AND\hat{\Pi}',
\]
yielding that $\PhiE{\hat{\Pi}} $
is equivalent to
\[
\hat{\Pi}'\AND \neg\exists V\Big[(V<V')\AND\bigwedge_{r\in\Pi}\big(\PhiT{B(r)}\IMPL\PhiT{H(r)}\big)\AND\hat{\Pi}'\Big].
\]
Since no  atom from $V$ occurs in $\hat{\Pi}'$, by applying Proposition~\ref{prop:quantifier-shift} repeatedly, the occurrence of $\hat{\Pi}'$ in the scope of $\exists V$ can be moved outside the quantifier, obtaining
\begin{equation}\label{eq:transf:2}
\hat{\Pi}'\AND \neg\Big[\exists V\Big((V<V')\AND\bigwedge_{r\in\Pi}\big(\PhiT{B(r)}\IMPL\PhiT{H(r)}\big)\Big)\AND\hat{\Pi}'\Big].
\end{equation}
Using De Morgan's law, (\ref{eq:transf:2}) can then be rewritten into 
\begin{equation}\label{eq:transf:1}
\hat{\Pi}'\AND \Big[\neg\exists V\Big((V<V')\AND\bigwedge_{r\in\Pi}\big(\PhiT{B(r)}\IMPL\PhiT{H(r)}\big)\Big)\OR\neg\hat{\Pi}'\Big].
\end{equation}
By the distributivity law, $\neg\hat{\Pi}'$ finally
gets absorbed by the first conjunct of (\ref{eq:transf:1}).
The result of this manipulation is $\PhiEn{\Pi}$.
\end{proof}

\begin{example}
For illustration, let us analyse the functioning of $\PhiEn{\cdot}$ applied to 
the program $\Pi=\{ p \LPif (q\AND r) \OR (\neg q \AND \neg s)\}$ from Example~\ref{ex:lp}.
Then, $\PhiEn{\Pi}$ is given by 
$$
\hat{\Pi}' \AND \neg \exists V \Big[ (V<V') \AND \Big( \big((q\AND r) \OR  (\neg q' \AND \neg s')\big) \IMPL p\Big)\Big],
$$
where
$\hat{\Pi} = \big( (q\AND r) \OR (\neg q \AND \neg s) \big) \IMPL p$,
$V=\{p,q,r,s\}$, and
\[
\begin{array}{r@{~}l}
(V<V')=&(p\IMPL p')\AND(q\IMPL q')\AND(r \IMPL r')\AND(s \IMPL s')\AND \\[1ex]
&\neg\big( (p'\IMPL p)\AND(q'\IMPL q)\AND(r' \IMPL r)\AND(s' \IMPL s) \big).
\end{array}
\]

First, let us verify that the interpretation $I'=\{p'\}$---which corresponds to the only stable model $\{p\}$ of $\Pi$---is a model of $\PhiEn{\Pi}$.
Clearly, $\hat{\Pi}'$ is true under $I'$, so it remains to check whether
\begin{equation}\label{eq:exphi}
\exists V \Big[ (V<V') \AND \Big(\big( (q\AND r) \OR  (\neg q' \AND \neg s')\big) \IMPL p\Big)\Big]
\end{equation}
is false under $I'$.
By the properties of $<$ and the semantics of the existential quantifier, we have
that~(\ref{eq:exphi}) is true iff there is some $J\subset I$ such
that  
\begin{equation}\label{eq:exphi2}
\big( (q\AND r) \OR  (\neg q' \AND \neg s')\big) \IMPL p
\end{equation}
is true under $J\cup I'$. 
Since $I=\{p\}$, the only proper subset of $I$ is the empty set, but~(\ref{eq:exphi2}) is false under $I'$, because the antecedent of~(\ref{eq:exphi2}) is true under  $I'$ whereas $p$ is false under $I'$.
Hence, (\ref{eq:exphi}) is false under $I'$, and consequently $I'$ is a model of $\PhiEn{\Pi}$.

To verify that no other $I\subseteq V$ is a model of $\PhiEn{\Pi}$, we first check that no $I'$ with
$\{p'\}\subset I'\subseteq \{p',q',r',s'\}$ is a model of $\PhiEn{\Pi}$.
Obviously, such an $I'$ satisfies $\hat{\Pi}'$, but it also satisfies~(\ref{eq:exphi}), since there is some $J\subset I$, viz.\ $J=\{p\}$, such that $J\cup I'$
is a model of both $(V<V')$ and~(\ref{eq:exphi2}).
So, $I'$ is then not 
a model of $\PhiEn{\Pi}$. 
Finally, observe that the remaining interpretations, $\emptyset$,
$\{q'\}$, $\{r'\}$, and $\{s'\}$, are not models 
of $\PhiEn{\Pi}$ because they are not models of $\hat{\Pi}'$. 
\end{example}

Similar to the case of equilibrium logic, from $\PhiEn{\cdot}$ we straightforwardly obtain encodings  for the basic reasoning tasks in the context of logic programs.

\begin{corollary} \label{cor:queries}
Let $\Pi$ be a logic program, $\varphi$ a formula, $V=\var{\Pi}$, and $W=\var{\Pi}\cup\var{\varphi}$.
Then, 

\begin{enumerate}
\item $\Pi$ has a stable model iff $\models \exists V'\, \PhiEn{\Pi}$,

\item $\Pi\brave\varphi$ iff $\models\exists W'\, \big( \PhiEn{\Pi} \AND \varphi'\big)$, and

\item $\Pi\skept\varphi$ iff $\models\forall W'\, \big( \PhiEn{\Pi} \IMPL \varphi'\big)$.

\end{enumerate}

\end{corollary}

In conclusion, let us mention that Theorem~\ref{thm:stable:models} generalises a similar QBF encoding put forth by \citeN{Egly00c} for the case of disjunctive logic programs.
To state this result, let us introduce the following customary notation:
Let $\Pi$ be a disjunctive logic program and $r\in\Pi$.
We write $B^+(r)$ to refer to a conjunction of all positive literals in $B(r)$, and $B^-(r)$ stands for a conjunction of all negative literals in $B(r)$.
Then, \citeN{Egly00c} use the following QBF to express the stable models of a given disjunctive logic program $\Pi$:
\[
\TLP{\Pi} =  
\hat{\Pi}
\AND \neg \exists V' \Big[ (V' < V) \AND \bigwedge_{r\in\Pi} \Big(\big(B^+(r') \AND B^-(r)\big) \IMPL H(r') \Big)\Big].
\]
It holds that $I$ is a stable model of $\Pi$ iff $I$ is a model of $\TLP{\Pi}$, for any $I\subseteq V$.

It is easily seen that, given a disjunctive logic program $\Pi$, the transformation
$\PhiEn{\Pi}$ coincides with $\TLP{\Pi}$, providing the priming of formulas is interchanged.

\subsection{Relation to Circumscription}\label{sec:circ}

By using the language of quantified propositional logic, we were able to conveniently express the inherent minimality postulates of both equilibrium models and stable models in terms of the operator $<$ and existential quantification.
In this section, we relate our encodings to circumscription~\cite{McCarthy80}, a well known technique to realise minimal-model reasoning.\footnote{Part of our discussion is also given by \citeN{Ferraris06}.}

For the case of disjunctive logic programs, an early result about the relation between stable models and circumscription was established by \citeN{Lin91}.
This result holds for nested logic programs and equilibrium logic as well, as discussed by \citeN{Ferraris:Lee:Lifschitz:07}, yet we provide independent proofs from our characterisations.

\medskip
Originally, as put forth by \citeN{McCarthy80}, circumscription is defined as a special formula schema of second-order logic.
In the propositional case, which is the relevant setting for our purposes here, circumscription is actually a formula of quantified propositional logic and can be defined in the following way (cf. also \citeN{Lifschitz94}).

Let $T$ be a theory and $(P,Q,Z)$ a partition of $\var{T}$. 
For two (classical) models, $I$ and $J$, of $T$, we define $I \le_{P;Z} J$  iff 
\begin{enumerate}
\item $(I \cap Q) = (J\cap Q)$, and
\item $(I \cap P) \subseteq (J\cap P)$.
\end{enumerate}
A model $I$ of $T$ is $(P;Z)$-\emph{minimal} iff no model $J$ of $T$ with 
$J \ne I$ 
satisfies $J \le_{P;Z} I$.

Informally, the partition $(P,Q,Z)$ can be interpreted as follows: The set $P$
contains the variables to be minimised, $Z$ are those variables that can vary
in minimising $P$, and the remaining variables $Q$ are fixed in minimising
$P$.

Let $T$ be a theory and $(P,Q,Z)$ a partition of $\var{T}$, 
where $P=\{p_1 ,\ldots, p_n\}$ and 
$Z=\{z_1 ,\ldots, z_m\}$. 
The set of $(P;Z)$-minimal models of 
$T$ is given by the {models} of the QBF 
\[
\Circ{T;P;Z} =
T \AND \neg \exists \tilde{P}\,\exists\tilde{Z} \Big( (\tilde{P} < P) \AND T[P/\tilde{P},Z/\tilde{Z}]  \Big),
\]
where $\tilde{P}=\{\tilde{p}_1 ,\ldots, \tilde{p}_n\}$ and $\tilde{Z}=\{\tilde{z}_1 ,\ldots, \tilde{z}_m\}$ are sets of new variables corresponding to $P$ and $Z$,
respectively.
In what follows, we write $\Circ{T;P}$ for $\Circ{T;P;\emptyset}$.

The next result paraphrases the characterisation of \citeN{Lin91}.

\begin{proposition}\label{prop:lin}
Let $\Pi$ be a disjunctive logic program, $V=\var{\Pi}$, and $I\subseteq V$.
Then, $I$ is a stable model of $\Pi$ iff $I\cup I'$ is a model of 
\begin{equation}\label{eq:circ:lin}
\bigwedge_{p\in V} (p \IFF p')
\AND
\Circ{\bigwedge_{r\in\Pi}\big((B^+(r)\AND B^-(r')) \IMPL H(r)\big);V}.
\end{equation}
\end{proposition}

Using our terminology, it is quite obvious that 
$\bigwedge_{r\in\Pi}\big((B^+(r)\AND B^-(r')) \IMPL H(r)\big)$
is given by $\bigwedge_{r\in\Pi}\big(\PhiT{B(r)} \IMPL \PhiT{H(r)}\big)$.
Furthermore, let us abbreviate the formula $(S\leq T)\AND(T\leq S)$ by  $S=T$, for all sets $S=\{\varphi_1\commadots\varphi_n\}$ and $T=\{\psi_1\commadots\psi_n\}$ of formulas.
So, $S=T$ is obviously equivalent to
$\bigwedge_{i=1}^n (\varphi_i \IFF \psi_i)$. 
Then, we can rewrite~(\ref{eq:circ:lin}) into the following QBF:
\[
(V=V') \AND \Circ{\bigwedge_{r\in\Pi}\big(\PhiT{B(r)} \IMPL \PhiT{H(r)}\big);V} .
\]

Now, this formula schema is directly applicable to arbitrary nested logic programs as well:

\begin{theorem}\label{thm:circ:nested}
Let $\Pi$ be a logic program, $V=\var{\Pi}$, and $I\subseteq V$.
Then, $I$ is a stable model of $\Pi$ iff $I\cup I'$ is a model of 
\begin{equation}\label{eq:circ:nested}
(V=V')
\AND
\Circ{\bigwedge_{r\in\Pi}\big(\PhiT{B(r)} \IMPL \PhiT{H(r)}\big);V} .
\end{equation}

\end{theorem}

\begin{proof}
In view of Theorem~\ref{thm:stable:models}, we know that $I$ is a stable model of $\Pi$ iff $I'$ is a model of
$$
\PhiEn{\Pi} =  
\hat{\Pi}'
\AND 
\neg \exists V \Big(
(V < V') \AND 
\bigwedge_{r \in \Pi} \!
\big(
\PhiT{B(r)} \IMPL \PhiT{H(r)} \big)
\Big).
$$
Now, clearly, $I'$ is a model of $\PhiEn{\Pi}$ iff $I\cup I'$ is a model of $(V=V')
\AND\PhiEn{\Pi}$.
It thus suffices to show that the latter formula is equivalent to (\ref{eq:circ:nested}). This can be seen as follows.

By the definition of circumscription, formula~(\ref{eq:circ:nested})
is the following QBF:
\begin{equation}\label{eq:circsm1}
(V=V')
\AND
\PhiTn{\Pi} \AND
\neg\exists\widetilde{V} 
\Big( (\widetilde{V} < V) \AND 
(\PhiTn{\Pi})\sub{V}{\widetilde{V}}
\Big),
\end{equation}
where $\PhiTn{\Pi}=\bigwedge_{r \in \Pi} \!
\big(
\PhiT{B(r)} \IMPL \PhiT{H(r)} \big)$.
Observe that the atoms from $V'$ are \emph{not} among the atoms 
renamed by
$(\PhiTn{\Pi})\sub{V}{\widetilde{V}}$.
Since furthermore the formula $(V=V')$
enforces that the same truth value is assigned to 
$p$ and $p'$, for each $p\in V$, 
QBF~(\ref{eq:circsm1}) is obviously equivalent to
\begin{equation}\label{eq:circsm:2}
(V=V')
\AND 
\PhiTn{\Pi} \AND
\neg\exists\widetilde{V}
\Big( (\widetilde{V} < V') \AND 
(\PhiTn{\Pi})\sub{V}{\widetilde{V}}
\Big),
\end{equation}
by using $V'$ instead of $V$ in the second argument of $<$.
Now, in the presence of $(V=V')$,
$\PhiTn{\Pi}$ is equivalent to $\hat{\Pi}'$.
Furthermore, no atom from $V$ occurs in 
$$
\neg \exists \widetilde{V}
\Big( (\widetilde{V} < V') \AND
(\PhiTn{\Pi})\sub{V}{\widetilde{V}}
\Big).
$$
Therefore, we can replace the existentially quantified variables $\widetilde{V}$ by $V$, 
from which we obtain that (\ref{eq:circsm:2}) is equivalent to
$$
(V=V')
\AND \hat{\Pi}' \AND
\neg \exists V\Big( (V < V') \AND  \PhiTn{\Pi} \Big),
$$
which is $(V=V')\AND\PhiEn{\Pi}$.
We thus showed that the latter formula is indeed equivalent to~(\ref{eq:circ:nested}). 
\end{proof}

Even more generally, Theorem~\ref{thm:circ:nested} can be extended to full equilibrium logic in the following fashion:

\begin{theorem}\label{thm:circ:eql}
Let $\varphi$ be a formula, $V=\var{\varphi}$, and $I\subseteq V$.
Then, $\tuple{I,I}$ is an equilibrium model of $\varphi$ iff $I\cup I'$ is a model of 
\begin{equation}\label{eq:circ:eql}
(V=V')
\AND
\Circ{\PhiT{\varphi};V}.
\end{equation}

\end{theorem}

\begin{proof}
The proof proceeds along the same line of reasoning as the proof of Theorem~\ref{thm:circ:nested}, except by resorting to Theorem~\ref{thm:eqm} and by showing that 
$(V=V')
\AND\PhiE{\Pi}$ is equivalent to~(\ref{eq:circ:eql}).
\end{proof}

As noted above, these generalisations of Lin's result have also been discussed by \citeN{Ferraris:Lee:Lifschitz:07},\footnote{In fact, since they treat the stable models of first-order formulas, their work is situated within full second-order logic rather than quantified propositional logic.} but they use an unoptimised version of  $\PhiT{\cdot}$ where a negated formula $\neg\psi$ is mapped to $ \neg\PhiT{\psi}\AND\neg\psi'$ (recall our discussion about this point in Section~\ref{sec:eqlogic-char}), so our results provide more compact encodings.

\medskip
For the case of disjunctive programs without negation, it is well known \cite{bara-2002} that the stable models of a program coincide with its minimal models.
We can extend this observation to nested logic programs as follows:
Let us call a program \emph{positive} if the expressions in 
the bodies and heads of its rules do not contain any negation.
Now, for any expression $\varphi$ without negation, it holds that $\PhiT{\varphi}=\varphi$.
Hence, for any positive program $\Pi$ with $\var{\Pi}=V$,
the encoding~(\ref{eq:circ:nested}) from Theorem~\ref{thm:circ:nested} 
reduces to 
\[
(V=V') \AND \Circ{ \bigwedge_{r\in \Pi} \big( B(r)\IMPL H(r)\big);V},
\]
which is just
\[
(V=V') \AND \Circ{\hat{\Pi};V}.
\]
Since no primed variables occur in $\Circ{\hat{\Pi};V}$ anymore,
we can safely drop the first conjunct $(V=V')$ and get as result
a purely circumscriptive theory, with all atoms being minimised.
We can thus state:

\begin{theorem}
For any positive program $\Pi$ with $\var{\Pi}=V$, the stable models of $\Pi$ are
given by $\Circ{\hat{\Pi};V}$.
\end{theorem}

Rounding out our exposition about circumscription, let us mention that, prior to \citeN{Lin91}, the relation between default negation and circumscription was already investigated by Gelfond et al.\
\shortcite{Gelfond89,Gelfond90}, showing that the stable models of programs with
a restricted use of negation (viz.\ with \emph{stratified negation}) can be characterised
via the minimal models of a more involved method of circumscription, called
\emph{iterated circumscription}.
Incidentally, a similar result, albeit
in terms of \emph{prioritised circumscription}, is given by \citeN{bara-2002}.

Yet another circumscription technique closely related to stable models is \emph{introspective circumscription}, due to \citeN{Lifschitz89}.
This kind of circumscription is a further development of \emph{autocircumscription}~\cite{Perlis88}, making use of an epistemic operator $L$, where $L\varphi$ informally means that an agent believes $\varphi$.
As shown by \citeN{gelfond-etal94}, introspective circumscription is (in the finite propositional case) basically equivalent to autoepistemic logic~\cite{Moore85}.
Actually, the latter is a very expressive nonmonotonic logic that can host a variety of different formalisms, just like quantified propositional logic.\footnote{See, \egc \citeN{yuan94} for a study on the expressive power of autoepistemic logic.}
In fact, the stable-model semantics for logic programs was originally developed having the epistemic approach of logic programming in mind, where default negation is interpreted as the combination $\neg L$ in autoepistemic logic, as first observed by \citeN{Gelfond87}.
In subsequent research, various embeddings of logic programs under the stable-model semantics (for different syntactic fragments) into autoepistemic logic have been discussed~\cite{marek93:_reflex_autoep_logic_logic_progr,lifschitz93:_exten,chen93:_minim,przymusinski91:_stabl_seman_disjun_progr,bruijn07-embed-non-groun-logic-progr}.

In what follows, we briefly indicate the relation between stable models of logic programs and propositional introspective circumscription for the case of normal logic programs, as presented by \citeN{Lifschitz89}.

The language of propositional introspective circumscription assumes standard propositional variables as well as \emph{modal atoms} of form $L p$, where $p$ is a propositional variable.
Formulas are then built as combinations of propositional variables and modal atoms using the sentential connectives $\neg$, $\vee$, $\wedge$, and 
$\IMPL$ as usual.
The modal atoms are treated semantically like propositional variables, just forming a disjoint class of atoms like the primed atoms in our notation.
Hence, the semantics of quantified propositional logic is directly extended to this class of formulas having access to an additional set of atoms.
Given a set $V$ of propositional variables, let us write $LV$ to refer to the set $\{Lp\mid p\in V\}$.

Let $T$ be a theory built from a set $V$ of propositional variables and a set $M$ of modal atoms.
Then, the \emph{introspective circumscription} of $T$ is given by the formula 
$$
\ICirc{T} = T \AND \bigwedge_{Lp \in M\cup LV} \Big( Lp \IFF 
\big( \forall \tilde{V} ( T[V/\tilde{V}] \IMPL \tilde{p})\big)\Big),
$$
where $\tilde{V}$ is a set of new variables corresponding to the propositional variables occurring in $T$.
Furthermore, for a normal logic program $\Pi$, let $\mu(\Pi)$ be the theory consisting of all formulas 
$$
p_1 \AND \cdots \AND p_m \AND \neg L q_1 \AND \cdots \AND \neg L q_n \IMPL h,
$$
where
$
h \LPif p_1 \AND \cdots \AND p_m \AND \neg q_1 \AND \cdots \AND \neg q_n 
$
is in $\Pi$.
Then, the following result expresses the relation shown by \citeN{Lifschitz89}:
\begin{proposition}
Let $\Pi$ be a normal logic program, $V=\var{\Pi}$, and $I \subseteq V$.
Then, $I$ is a stable model of $\Pi$ iff 
$LI$ is a model of 
\begin{equation}\label{eq:icirc}
\forall V\big(\mu(\Pi)\IMPL \ICirc {\mu(\Pi)}\big).
\end{equation}
\end{proposition}
\begin{example}
For illustration, consider the program $\Pi=\{p\LPif \neg q\}$.
Then, $\mu(\Pi)=\{\neg Lq\IMPL p\}$, and $\ICirc{\mu(\Pi)}$ is given by the formula
$$
(\neg Lq\IMPL p)\AND \Big(Lp\equiv \forall\tilde{p}\forall\tilde{q}\big((\neg Lq\IMPL \tilde{p})\IMPL \tilde{p} \big) \Big) \AND \Big(Lq\equiv \forall\tilde{p}\forall\tilde{q} \big((\neg Lq\IMPL \tilde{p})\IMPL \tilde{q} \big) \Big).
$$
It can be shown that this formula is equivalent to
$$
(\neg Lq\IMPL p)\AND (Lp\equiv\neg Lq)\AND \neg Lq,
$$
which in turn reduces to
$$
p\AND Lp \AND \neg Lq.
$$
Hence, (\ref{eq:icirc})
is equivalent to $$\forall p \big((\neg Lq\IMPL p)\IMPL (p\AND Lp \AND \neg Lq)\big).$$ 
The last formula has just one model, viz.\ $\{Lp\}$, which corresponds to the single stable model $\{p\}$ of $\Pi$.
\end{example}

\subsection{Characterising Different Notions of Equivalence}\label{sec:eq}

We now turn our attention to different notions of
equivalence which have been studied in the context of equilibrium logic and nested logic programs.
In the same fashion as before, we first deal with the general case---that is to say, with theories in equilibrium logic---and afterwards, as a special case, with nested logic programs.

In classical logic, the well known \emph{replacement property} holds, according to which any formula $\varphi$ occurring as a specified part in a formula $C_\varphi$ can be replaced by any logically equivalent formula $\psi$ yielding a formula $C_\psi$ which is still logically equivalent to $C_\varphi$.\footnote{The replacement property is also known as substitution \emph{salva veritate} in view of Leibniz's principle ``eadem sunt, quorum unum potest substitui alteri salva veritate''.}
In a nonmonotonic setting, however, when interpreting logical equivalence in the traditional way as the relation which holds in case two theories have the same ``intended models''(like, two theories in equilibrium logic have the same equilibrium models or two logic programs have the same 
stable models), such a property fails in general.
Thus, in order to ensure the validity of a replacement property in nonmonotonic formalisms, more robust notions of equivalence have to be considered.
In fact, a suitable notion to that effect is \emph{strong equivalence}, first introduced and studied by \citeN{Lifschitz01} for nested logic programs and equilibrium logic.
A weaker variant of strong equivalence is \emph{uniform equivalence}, first discussed by \citeN{Eiter03} for disjunctive logic programs and subsequently extended by \citeN{Pearce04} for nested programs and equilibrium logic.

In what follows, we characterise strong and uniform equivalence, along with the ``traditional'' concept of equivalence, in terms of quantified propositional logic.
In fact, strong equivalence will be captured by means of ordinary classical logic, \iec without requiring any quantifier.  

We start with formally defining the equivalence relations under consideration.
Let $T_1$ and $T_2$ be theories.
Then,
\begin{enumerate} 
\item
$T_1$ and $T_2$ are (\emph{ordinarily}) \emph{equivalent},
in symbols 
$T_1\equiv_o T_2$,
iff they possess the same equilibrium models,

\item
$T_1$ and $T_2$ are \emph{uniformly equivalent},
in symbols $T_1\equiv_u T_2$, iff,
for every set $R$ of atoms,
$T_1\cup R\equiv_o T_2\cup R$, and

\item
$T_1$ and $T_2$ 
are \emph{strongly equivalent}, 
in symbols $T_1 \equiv_s T_2$, iff,
for every theory $S$, $T_1 \cup S\equiv_o T_2\cup S$.
\end{enumerate}

In case of programs, analogous relations are defined as follows:
For all programs $\Pi_1$ and $\Pi_2$,
\begin{enumerate} 
\item
$\Pi_1$ and $\Pi_2$ are (\emph{ordinarily}) \emph{equivalent},
in symbols 
$\Pi_1\equiv_o \Pi_2$,
iff they possess the same stable models,

\item
$\Pi_1$ and $\Pi_2$ are \emph{uniformly equivalent},
in symbols $\Pi_1\equiv_u\Pi_2$, iff,
for every program $\Pi$ containing facts only, $\Pi_1\cup\Pi\equiv_o\Pi_2\cup \Pi$,
and

\item
$\Pi_1$ and $\Pi_2$ 
are \emph{strongly equivalent}, 
in symbols $\Pi_1\equiv_s\Pi_2$, iff, 
for every program $\Pi$,
$\Pi_1 \cup \Pi\equiv_o \Pi_2\cup\Pi$.
\end{enumerate}
Taking the association between programs and theories into account, it can easily be seen that, for every program $\Pi_1$ and $\Pi_2$, we actually have $\Pi_1\equiv_e\Pi_2$ iff $\hat{\Pi}_1\equiv_e\hat{\Pi}_2$, for $e\in\{o,s,u\}$.

Clearly, strong equivalence implies uniform equivalence, which in turn implies ordinary equivalence. 
However, it is a straightforward matter to verify that the implications in the other direction fail and therefore each notion is distinct. 
This already holds for the special case of logic programs, 
as shown by the following example.

\begin{example}\label{ex:lp2}

Consider the
programs $\{p\LPif\}$ and $\{p\LPif \neg q\}$, 
which are easily verified to be ordinarily equivalent but not uniformly so (just add the fact $q\LPif$ to each program).
Similarly, strong and uniform equivalence are distinct in view of the programs 
$$
\Pi_1=\{ p\OR q \LPif\, \}\quad\mbox{and}\quad
\Pi_2=\{p\LPif\neg q,\,\, q\LPif \neg p\}.
$$
Indeed, both programs have  $\{p\}$ and $\{q\}$ as their stable models, so
they are ordinarily equivalent. 
But
$\Pi_1$ and $\Pi_2$ are not strongly equivalent, since adding
rules 
$$
\{p\LPif q,\,\, q\LPif p\}
$$
to $\Pi_1$ yields a program whose only stable model is $\{p,q\}$, whereas adding the same rules to $\Pi_2$ results in a program having  no stable model at all.
However, $\Pi_1$ and $\Pi_2$ are uniformly equivalent, as can be seen as follows.
Adding the fact $p\LPif$ to both programs leaves the resultant programs equivalent,
with $\{p\}$ as their single stable model. 
Analogously, equivalence
is preserved when fact $q\LPif$ is added; and, similarly,
if both $p\LPif$ and $q\LPif$
are added as new facts. In the latter case, both extended programs have
the single stable model $\{p,q\}$.
Clearly, adding any other fact just adds the respective atom to the
stable models of both programs.
\end{example}

In what follows, we recall some important characterisations forming
the basis for our subsequent encodings of strong and uniform equivalence, respectively.
To begin with, the following property is the central result of \citeN{Lifschitz01}:

\begin{proposition} \label{prop:strong}
Two theories 
are strongly equivalent iff
they are equivalent in the logic of here-and-there.
\end{proposition}

It is worth noting that \citeN{Jongh02} showed that the weakest intermediate logic that can replace 
here-and-there in the 
characterisation of
strong equivalence for nested logic programs 
is the  logic KC of weak excluded
middle~\cite{Kowalski68}. 
The logic KC is axiomatised by intuitionistic logic together with
the schema
$\neg \varphi \OR \neg \neg \varphi$. 
Since HT is actually the 
greatest super-intuitionistic logic capturing strong equivalence, 
the results by Hendriks and de~Jongh show that all and only 
the super-intuitionistic logics lying between KC and HT capture 
strong equivalence for nested programs. 

Proposition~\ref{prop:strong} 
was also reformulated for the case of logic programs by Turner \shortcite{Turner01,Turner03},
replacing equivalence in the logic HT by the condition that two programs have the same \emph{SE-models}.
The latter are, like HT-models, ordered pairs of interpretations, but they are defined in terms of the reduct of a program instead of referring to HT explicitly.
In fact, as shown by \citeN{Turner03}, HT-models and SE-models are identical concepts
for the programs under consideration (\iec for
programs without strong negation).

In subsequent work,  \citeN{Eiter03}
used SE-models to characterise uniform equivalence 
for disjunctive logic programs. 
They provided a general condition which holds also for the case of infinite programs and introduced \emph{UE-models} to characterise uniform equivalence between finite programs.
Roughly speaking, UE-models are SE-models which are maximal elements under a certain ordering over SE-models.
Although \citeN{Eiter03} mentioned that their characterisation can be extended for nested logic programs as well, no formal account was provided to that effect.
However, this point was addressed in subsequent work by \citeN{Pearce04}, extending the characterisations of \citeN{Eiter03} to full equilibrium logic.
In order to attain sufficient generality for our purposes, we will therefore base our encodings for testing uniform equivalence on the exposition provided by \citeN{Pearce04}.
In what follows, we first present the relevant property of the latter work, and afterwards we reformulate their characterisation to better suit our needs.

\begin{proposition}\label{prop:ue}
Let $T_1$ and $T_2$ be two theories which are classically equivalent.
Then, 
$T_1$ and $T_2$ are 
uniformly equivalent iff, for every $I_H,I_T\subseteq\var{T_1\cup T_2}$, the following two conditions jointly hold:
\begin{itemize}
\item[(i)] If $\tuple{I_H,I_T}$ is a non-total HT-model of $T_1$, then there is an interpretation $J$ such that $I_H\subseteq J \subset I_T$
and $\tuple{J,I_T}$ is an HT-model of $T_2$.

\item[(ii)]
If $\tuple{I_H,I_T}$ is a non-total HT-model of $T_2$, then there is an interpretation $J$ such that $I_H\subseteq J \subset I_T$
and $\tuple{J,I_T}$ is an HT-model of $T_1$.
\end{itemize}

\end{proposition}
Let us introduce the following notation: 
For all theories $T_1$ and $T_2$, define $U(T_1,T_2)$ as the set of all pairs $\tuple{I_H,I_T}$, where
$I_H,I_T\subseteq\var{T_1\cup T_2}$, such that whenever $\tuple{I_H,I_T}$ is a non-total HT-model of $T_1$, then there is an interpretation $J$ such that $I_H\subseteq J \subset I_T$
and $\tuple{J,I_T}$ is an HT-model of $T_2$.
Clearly, $\tuple{I_H,I_T}\in U(T_1,T_2)$ iff $\tuple{I_H,I_T}$ satisfies Condition~(i) of the above proposition and, symmetrically, $\tuple{I_H,I_T}\in U(T_2,T_1)$ iff $\tuple{I_H,I_T}$ satisfies Condition~(ii).
Hence, Proposition~\ref{prop:ue} can be rephrased thus:

\begin{corollary}\label{cor:ue}
Two theories $T_1$ and $T_2$ are 
uniformly equivalent iff
\begin{itemize}
\item[(i)] $T_1$ and $T_2$
are equivalent in classical logic and

\item[(ii)]
$\tuple{I_H,I_T}\in U(T_1,T_2)\cap U(T_2,T_1)$, for every $\tuple{I_H,I_T}$ with $I_H,I_T\subseteq\var{T_1\cup T_2}$. 
\end{itemize}
\end{corollary}

We now describe the encodings of the equivalence notions under consideration in terms of quantified propositional logic.
To begin with, we give a characterisation of the set $U(T_1,T_2)$, for every theory $T_1$ and $T_2$, used for the encoding of checking uniform equivalence.

In what follows, we denote by $\PhiTpp{\cdot}$ the mapping obtained from the translation $\PhiT{\cdot}$ of Definition~\ref{def:encoding} by replacing each unprimed atom $v$ in $\PhiT{\cdot}$ by a globally new atom $v''$.  
Recall that, as per our priming convention, for every set $V$ of atoms, $V'$ and $V''$ are defined as the sets $\{v'\mid v\in V\}$ and $\{v''\mid v\in V\}$, respectively, being disjoint from $V$. 

\begin{lemma} \label{lemma:preceq}
Let $T_1$ and $T_2$ be theories, $V=\var{T_1\cup T_2}$, and $I_H,I_T\subseteq V$ interpretations.
Then, $\tuple{I_H,I_T}\in U(T_1,T_2)$ iff
$I_H \cup I'_T$ is a model of 
\[
\eqmod{T_1}{T_2} = 
\big((V < V') \AND \PhiT{T_1}\big) \IMPL \exists V'' \big( (V\leq V'') \AND (V''< V') \AND \PhiTpp{T_2}\big).
\]
\end{lemma}

\begin{proof}
We show the following two properties:

\begin{description}

\item[\rm ($\alpha$)] $\tuple{I_H,I_T}$ is a non-total HT-model of $T_1$ iff $I_H\cup I_T'$ is a model of
\[
\varphi=(V < V') \AND \PhiT{T_1}.
\]

\item[\rm ($\beta$)] $\tuple{J,I_T}$ is an HT-model of $T_2$ with $I_H\subseteq J\subset I_T$ iff $I_H\cup I_T'\cup J''$ is a model of
\[
\psi=(V\leq V'') \AND (V''< V') \AND \PhiTpp{T_2}.
\]
\end{description}
From this, by the semantics of the conditional $\IMPL$ and of existential quantification, as well as by the definition of $U(T_1,T_2)$, the claim of the lemma is an immediate consequence.

We start by proving Property~($\alpha$).
By Lemma~\ref{thm:HTSat}, we have that $\tuple{I_H,I_T}$ is an HT-model of $T_1$ iff $I_H \cup I'_T$ is a model of $(V\leq V')\AND\PhiT{T_1}$.
Moreover, $\tuple{I_H,I_T}$ is non-total iff $I_H\subset I_T$.
So, by the semantics of $<$, $\tuple{I_H,I_T}$ is non-total iff $I_H \cup I'_T$ is a model of $(V\leq V')\AND\neg (V'\leq V)$.
It follows that $\tuple{I_H,I_T}$ is a non-total HT-model of $T_1$ iff $I_H \cup I'_T$ is a model of 
$(V<V')\AND\PhiT{T_1}$.
Thus, Property~($\alpha$) holds.

We continue with the proof of Property~($\beta$).
By Property~($\alpha$), it holds that $\tuple{J,I_T}$ is an HT-model of $T_2$ such that $J\subset I_T$ iff $J\cup I_T'$ is a model of $(V<V')\AND\PhiT{T_2}$.
By a simple renaming we get that the latter holds iff $J''\cup I_T'$ is a model of $(V''<V')\AND\PhiTpp{T_2}$.
Furthermore, the semantics of $<$ tells us that $I_H\subseteq J$ iff $I_H\cup J'$ is a model of $V\leq V'$.
Applying again a simple renaming, we get that $I_H\subseteq J$ iff $I_H\cup J''$ is a model of $V\leq V''$.
Combining these conditions, we establish that $\tuple{J,I_T}$ is an HT-model of $T_2$ satisfying $I_H\subseteq J\subset I_T$ precisely when 
(i)~$I_H\cup J''$ is a model of $V\leq V''$ and 
(ii)~$J''\cup I_T'$ is a model of $(V''<V')\AND\PhiTpp{T_2}$.
But $V\leq V''$ contains no atoms from $V'$ and $(V''<V')\AND\PhiTpp{T_2}$ contains no atoms from $V$, so, since $V$, $V'$, and $V''$ are pairwise distinct, it follows that 
$\tuple{J,I_T}$ is an HT-model of $T_2$ satisfying $I_H\subseteq J\subset I_T$ iff
$I_H\cup I_T'\cup J''$ is a model of 
\[
(V\leq V'')\AND(V''<V')\AND\PhiTpp{T_2},
\]
which proves Property~($\beta$).
\end{proof}

We are now in a position to state the main result of this section:

\begin{theorem}\label{thm:eqall}
Let $T_1$ and $T_2$ be theories and $V=\var{T_1\cup T_2}$. Then,
\begin{enumerate}
\item[(i)]
$T_1 \equiv_o T_2$ iff 
$
\models \forall V' (\PhiE{T_1}  \IFF  \PhiE{T_2})$, 
\item[(ii)]
$T_1 \equiv_u T_2$ iff 
$
\models \forall V \forall V' \big( (T_1 \IFF T_2) \AND \eqmod{T_1}{T_2} \AND \eqmod{T_2}{T_1}\big)$, and 
\item[(iii)]
$T_1 \equiv_s T_2$ iff 
$
\models \forall V \forall V' \big( (V\leq V') \IMPL (\PhiT{T_1}\IFF \PhiT{T_2})\big)$.
\end{enumerate}
\end{theorem}

\begin{proof}
Concerning Part~(i), from Theorem~\ref{thm:eqm} we know that for every $I\subseteq V$, 
$\langle I,I\rangle$ is an equilibrium model of $T_i$ iff $I'$ is a model of
$\PhiE{T_i}$, for $i\in\{1,2\}$. 
Hence, $T_1$ and $T_2$ possess the same equilibrium models, \iec 
$T_1\equiv_o T_2$ holds, iff $\forall V' (\PhiE{T_1}  \IFF  \PhiE{T_2})$ is valid.

Now consider Part~(ii).
By Corollary~\ref{cor:ue}, $T_1$ and $T_2$ are uniformly equivalent iff (a)~$T_1$ and $T_2$ are equivalent in classical logic and (b)~$\tuple{I_H,I_T}\in U(T_1,T_2)\cap U(T_2,T_1)$, for every pair $\tuple{I_H,I_T}$ with $I_H,I_T\subseteq \var{T_1\cup T_2}$.
Condition~(a) is clearly equivalent to the fact that $\models \forall V\forall V'(T_1\IFF T_2)$, and, by Lemma~\ref{lemma:preceq}, Condition~(b) holds precisely in the case that $\models \forall V\forall V'(\eqmod{T_1}{T_2} \AND \eqmod{T_2}{T_1})$.
So, (a) and (b) jointly hold iff
\begin{equation}\label{eq:equiv-main:1}
\forall V\forall V'(T_1\IFF T_2)\AND\forall V\forall V'(\eqmod{T_1}{T_2} \AND \eqmod{T_2}{T_1})
\end{equation}
is valid.
But (\ref{eq:equiv-main:1}) is easily seen to be equivalent to
\[
\forall V\forall V'\big((T_1\IFF T_2)\AND\eqmod{T_1}{T_2} \AND \eqmod{T_2}{T_1}\big),
\]
which establishes Part~(ii).

It remains to show Part~(iii).
By Proposition~\ref{prop:strong}, $T_1$ and $T_2$ are strongly equivalent iff 
the HT-models of $T_1$ and $T_2$ coincide. By Lemma~\ref{thm:HTSat}, 
the latter is the case iff $\PhiHT{T_1}$ and $\PhiHT{T_2}$ are logically equivalent
in classical logic, which in turn is equivalent to the condition that
\begin{equation}\label{eq:equiv-main:2}
\forall V V' \big( \PhiHT{T_1} \IFF \PhiHT{T_2}\big)
\end{equation}
is valid in quantified propositional logic.
Now, by the definition of $\PhiHT{\cdot}$, Formula~(\ref{eq:equiv-main:2})
is given by
$$
\forall V V' \Big( \big((V'\leq V)\AND \PhiT{T_1}\big) \IFF  \big((V'\leq V)\AND\PhiT{T_2}\big)\Big).
$$
By simple manipulations in classical logic, the latter formula can be transformed in an equivalence-preserving way into
\[
\forall V \forall V' \big( (V\leq V') \IMPL (\PhiT{T_1}\IFF \PhiT{T_2})\big).
\]
Thus, Part~(iii) holds.
\end{proof}

Note that the logical complexities of all encodings from 
Theorem~\ref{thm:eqall} are at most super-linear in the logical complexities of 
the two theories compared.
Furthermore, the encodings for ordinary and uniform equivalence possess one quantifier alternation, with a leading prefix of universal quantifiers, while the encoding for strong equivalence
amounts to a validity test in classical propositional logic. 
We have more to say about the consequences of these properties in the next section.

Clearly, for comparing programs rather than theories, we can make use of the above encodings as well, just by taking $\hat{\Pi}$ for each program $\Pi$ as the respective argument.
However, for testing ordinary equivalence, we can obtain a more compact encoding by directly resorting to $\PhiEn{\cdot}$ instead of $\PhiE{\cdot}$.
The corollary below summarises these observations:

\begin{corollary}\label{cor:eqall}
Let $\Pi_1$ and $\Pi_2$ be logic programs and $V=\var{\Pi_1\cup\Pi_2}$. Then,
\begin{enumerate}
\item
$\Pi_1 \equiv_o \Pi_2$ iff 
$
\models \forall V' (\PhiEn{\Pi_1}  \IFF  \PhiEn{\Pi_2}),
$ 
\item 
$\Pi_1 \equiv_u \Pi_2$ iff 
$\models \forall V \forall V' \Big( (\hat{\Pi}_1 \IFF \hat{\Pi}_2) \AND \eqmod{\hat{\Pi}_1}{\hat{\Pi}_2} \AND \eqmod{\hat{\Pi}_2}{\hat{\Pi}_1}\Big)$, and 
\item 
$\Pi_1 \equiv_s \Pi_2$ iff 
$
\models \forall V \forall V' \Big( (V\leq V') \IMPL (\PhiT{\hat{\Pi}_1}\IFF \PhiT{\hat{\Pi}_2})\Big).
$
\end{enumerate}
\end{corollary}

We remark that the logical complexities of
all encodings in Corollary~\ref{cor:eqall} are linear in $\lc{\hat{\Pi}_1}$ and $\lc{\hat{\Pi}_2}$.
While this is obvious as far as the encoding for ordinary equivalence is concerned,
since we already remarked that $\lc{\PhiEn{\Pi}}$ is linear in $\lc{\hat{\Pi}}$, for every program $\Pi$, 
for the other encodings this follows again from the observation that $\hat{\Pi}$ does not involve nested implications.

Our above encoding for uniform equivalence is based on the characterisation given by 
Proposition~\ref{prop:ue}.
In previous work, \citeN{Woltran04} showed that checking uniform equivalence between disjunctive logic programs can be reduced to checking ordinary equivalence between those kinds of programs.
We now generalise this result to arbitrary theories of equilibrium logic, thus providing an alternative encoding for checking uniform equivalence.
We start with extending the relevant result from \citeN{Woltran04} to the case of arbitrary theories.
As a preparatory step, we give a slight paraphrase of a result due to \citeN{Ferraris05},
which in turn is a generalisation of the well-known \emph{Splitting-Set Theorem} for disjunctive logic programs~\cite{lifs-turn-94,Eiter97a} to the case of equilibrium logic.

\begin{proposition}[Splitting-Set Theorem for Equilibrium Logic]\label{prop:splitting}
Let $S$ and $R$ be two theories such that each variable from $\var{S}$ occurs 
only in antecedents of implications or negated in $R$. Then, $\tuple{I,I}$ is an equilibrium model of $S\cup R$ 
iff 
\begin{itemize}
\item[(i)]
$\tuple{I\cap \var{S},I\cap \var{S}}$ is an equilibrium model of $S$ and
\item[(ii)] $\tuple{I,I}$ is an equilibrium model of $(I \cap \var{S})\cup R$.
\end{itemize}
\end{proposition}

In what follows, for every set $V$ of variables, define $V^\circ=\{v^\circ\mid v\in V\}$ as a set of new variables corresponding to the variables in $V$.

\begin{lemma}\label{lemma:UEalt}
Let $T_1$ and $T_2$ be theories and $V=\var{T_1\cup T_2}$. 
Then, $T_1\equiv_u T_2$ iff
$T^\sharp_1\equiv_o T^\sharp_2$, where
\[
T^\sharp_i = T_i \cup \{ \neg\neg v^\circ \IMPL v^\circ,\,
v^\circ \IMPL v \mid v\in V\},
\]
for $i=1,2$.
\end{lemma}

\smallskip\noindent
\emph{Proof}\\
The proof relies on the Splitting-Set Theorem for equilibrium logic.
Define
\[
S = \{ \neg\neg v^\circ \IMPL v^\circ \mid v\in V \}\quad\mbox{and}
\quad
R_i = \{ v^\circ \IMPL v \mid v \in V\}\cup T_i,
\] 
for $i=1,2$. 
Clearly, we have that
$T^\sharp_i = S \cup R_i$, for $i=1,2$. 
Now, since all variables from $S$ occur in $R_1$ and $R_2$ only
in implications of the form $v^\circ \IMPL v$,
by the Splitting-Set Theorem we get the following property:
\begin{enumerate}

\item[($\alpha$)]
for every interpretation $I$,
$\tuple{I,I}$ is an equilibrium model of $T^\sharp_i$ iff $\tuple{I\cap V^\circ,I\cap V^\circ}$ is an equilibrium model of $S$ and $\tuple{I,I}$ is an equilibrium model of $(I\cap V^\circ)\cup R_i$, for $i=1,2$.

\end{enumerate}

Furthermore, we require the following two properties, which are easily verified:

\begin{enumerate}

\item[($\beta$)] The set of all equilibrium models of $S$ is given by $\{\tuple{J,J}\mid J\subseteq V^\circ\}$.

\item[($\gamma$)]
For any $F\subseteq V$ and $i=1,2$,
if $\tuple{I,I}$ is an equilibrium model of $T_i\cup F$, then 
$\tuple{I\cup F^\circ,I\cup F^\circ}$ is an equilibrium model of $R_i\cup F^\circ$, and if $\tuple{J,J}$ is an equilibrium model of $R_i\cup F^\circ$ then,
$\tuple{V\cap J,V\cap J}$ is an equilibrium model of $T_i\cup F$.

\end{enumerate}

We proceed with the proof of the main result.
By definition, $T_1\equiv_u T_2$ 
iff
\begin{equation}\label{cond:1}
\mbox{$T_1\cup F \equiv_o T_2\cup F$, for every set $F$ of atoms.}
\end{equation}
Obviously, (\ref{cond:1}) is equivalent to the condition that
\begin{equation}\label{cond:2}
\mbox{$T_1\cup F \equiv_o T_2\cup F$, for every $F\subseteq V$.}
\end{equation}
However, Condition~($\gamma$) implies that, for every $F\subseteq V$,
$T_1\cup F \equiv_o T_2\cup F$
iff
$R_1\cup F^\circ\equiv_oR_2\cup F^\circ$.
Hence, (\ref{cond:2}) is equivalent to
\begin{equation}\label{cond:4}
\mbox{$R_1\cup F^\circ\equiv_oR_2\cup F^\circ$, for every $F\subseteq V$.}
\end{equation}
Now, in view of Conditions~($\alpha$) and ($\beta$), it is a straightforward matter to check that~(\ref{cond:4}) holds precisely in case that
$R_1 \cup S \equiv_o R_2\cup S$.
But the latter relation just states that $T^\sharp_1\equiv_o T^\sharp_2$.
So, in summary, the above chain of equivalences shows that $T_1\equiv_u T_2$ iff $T^\sharp_1\equiv_o T^\sharp_2$.
\hspace*{1em}\hbox{\proofbox}

\medskip

Exploiting the encoding of ordinary equivalence, as given by Theorem~\ref{thm:eqall},
we arrive at the following characterisation:

\begin{theorem}\label{cor:UEalt}
Let $T_1$ and $T_2$ be theories and $V=\var{T_1\cup T_2}$.
Furthermore, let ${T}^\sharp_1$ and ${T}^\sharp_2$ be defined as in Lemma~\ref{lemma:UEalt}.
Then,
$T_1 \equiv_u T_2$ iff 
$\models \forall V' (\PhiE{{T}^\sharp_1}  \IFF  \PhiE{{T}^\sharp_2})$.
\end{theorem}

A similar result can be shown for logic programs; we omit the obvious details.

In concluding this section, we compare our characterisation for testing strong equivalence with one by \citeN{Lin02}, which is devised for disjunctive logic programs, reducing the test of strong equivalence to checking entailment in classical propositional logic.\footnote{Lin's result was developed independently from our previous preliminary report~\cite{Pearce01}, where the characterisation for strong equivalence between theories of equilibrium logic, or between 
nested logic programs, respectively, 
in terms of classical logic was first reported. Also, Lin erroneously states in his discussion of our characterisation that we do not handle disjunctions, which we do, however.}
More specifically, Lin assigns to each disjunctive logic program $\Pi$ a set $\Gamma(\Pi)$ of formulas of classical propositional logic containing for each $r\in\Pi$ the two formulas
\[
(B^+(r)\AND B^-(r'))\IMPL H(r)\quad\mbox{and}\quad
(B^+(r')\AND B^-(r'))\IMPL H(r').
\]
Then, two programs, $\Pi_1$ and $\Pi_2$, with $V=\var{\Pi_1\cup \Pi_2}$, are strongly equivalent iff the following two assertions hold:
\begin{equation}
\{v\IMPL v'\mid v\in V\}\cup \Gamma(\Pi_1)\models \Gamma(\Pi_2),\label{lin:1}
\end{equation}
\begin{equation}
\{v\IMPL v'\mid v\in V\}\cup \Gamma(\Pi_2)\models \Gamma(\Pi_1).\label{lin:2}
\end{equation}
In terms of our notation, (\ref{lin:1}) and (\ref{lin:2}) can be restated thus:
\begin{equation}
\{(V\leq V') \AND \PhiT{\hat{\Pi}_1}\}\models \PhiT{\hat{\Pi}_2},\label{lin:3}
\end{equation}
\begin{equation}
 \{(V\leq V') \AND \PhiT{\hat{\Pi}_2}\}\models \PhiT{\hat{\Pi}_1}.\label{lin:4}
\end{equation}
But these two assertions are just reformulations of our characterisation from Corollary~\ref{cor:eqall}.
Indeed, by applying the deduction theorem from classical logic and some simple manipulations, (\ref{lin:3}) and (\ref{lin:4}) are equivalent to 
\[
\models (V\leq V') \IMPL \PhiT{\hat{\Pi}_1}\IFF \PhiT{\hat{\Pi}_2}.
\]
Adjoining universal quantifications for all variables in the above formula yields our characterisation.
Hence, our encoding directly generalises Lin's method.\footnote{We note that \citeN{Lin02} discusses also the case of disjunctive logic programs with variables; here, however, we do not consider theories or programs with variables.}

\section{Complexity}\label{sec:complexity}

In the previous section, we discussed encodings of the basic reasoning tasks associated with equilibrium logic and nested logic programs, viz.\ we dealt with the consistency problem as well as with brave and skeptical reasoning for both languages.
Additionally, we also captured the problem of checking ordinary, strong, and uniform equivalence between theories or programs.
Now, we analyse the computational complexity of these tasks.

A particular advantage of our encoding technique is that the quantifier structure of our translations yields in a direct manner upper complexity bounds for the corresponding problems.
This follows by invoking well-known complexity results about QBFs and by observing that our translations are constructible in polynomial time.
Moreover, for each of the upper bounds obtained in this fashion, we also show
that they are \emph{strict}, \iec they possess a matching 
lower bound. 
The results presented here generalise well-known complexity results for disjunctive logic programs under the stable-model semantics~\cite{Eiter95,Eiter03,Lin02}.

In what follows, we assume that the reader is familiar with the basic
concepts of complexity theory (cf., \egc \citeN{pap94} for a
comprehensive treatise on this subject).
For convenience, we briefly recapitulate the definitions and
some elementary properties of the complexity classes
considered in our analysis.
As usual, for any complexity class $C$, by $\co C$ we
understand the class of all problems which are
complementary to the problems in $C$.

Four complexity classes are relevant here, viz.\ $\NP$,  $\CONP$, $\SigmaP{2}$, and $\PiP{2}$.
In detail, the class $\NP$ consists of all decision problems
which can be solved with a nondeterministic Turing machine
working in polynomial time;
$\SigmaP{2}$ is the class of all problems solvable with a
nondeterministic Turing machine working in polynomial time having
access to an oracle for problems in $\NP$; and $\PiP{2}=\co\SigmaP{2}$.

Observe that the above classes are part of the \emph{polynomial hierarchy}~\cite{Stockmeyer76}, which is given by the following sequence of classes:
The initial elements are
\[\DeltaP{0}=\SigmaP{0}=\PiP{0}={\rm P},
\]
and, for $i>0$,
\[
\DeltaP{i} = \Pol^{\SigmaP{i-1}}, \mbox{ }
\SigmaP{i} = \NP^{\SigmaP{i-1}}\mbox{,   and }
\PiP{i}    = \co\NP^{\SigmaP{i-1}}.
\]
Here, $\Pol$ is the class of all problems solvable with a
deterministic Turing machine working in polynomial time, and, for
complexity classes $C$ and $A$, $C^A$ stands for
the \emph{relativised version} of $C$, \iec consisting of all problems which
can be decided by Turing machines of the same sort and time bound
as in $C$, only that the machines have access to an oracle for problems in $A$.
It holds that $\SigmaP{1}=\NP$, $\SigmaP{2}=\NP^\NP$, and $\PiP{2}=\co\NP^\NP$.
A problem is said to be at the \emph{$n$-th level} of the
polynomial hierarchy iff it is in $\DeltaP{n+1}$ and either
$\SigmaP{n}$-hard or $\PiP{n}$-hard.

The next proposition describes the close relation between the complexity classes $\SigmaP{n}$ and $\PiP{n}$, for $n\geq 1$, and QBFs having $n-1$ quantifier alternations.
Preparatorily, we introduce some further notation.

Let $\Phi$ be a closed prenex QBF of form 
$\quantifier_1 V_1  \quantifier_2 V_2 \ldots \quantifier_n V_n\, \phi$,
where
$\phi$ is a  propositional formula,
$\quantifier_i \in \{\exists,\forall\}$ such that $\quantifier_i\neq\quantifier_{i+1}$  
for $1\leq i \leq n-1$, and
$V_1\commadots V_n$ are pairwise disjoint sets
of propositional variables.
We call $\Phi$ an \emph{$(n,\exists)$-QBF} if
$\quantifier_1=\exists$, and an \emph{$(n,\forall)$-QBF} if
$\quantifier_1=\forall$.
Furthermore, we refer to $\phi$ as the \emph{matrix} of~$\Phi$.

\begin{proposition}[\citeNP{Wrathall76}]\label{prop:qsat}
For every $n\geq 0$, the following properties hold:
\begin{enumerate}
\item 
Deciding whether a given $(n,\exists)$-QBF $\Phi$ is valid is $\SigmaP{n}$-complete.
The problem remains $\SigmaP{n}$-hard even if the matrix of $\Phi$ is in conjunctive normal form and $n$ is odd,
or in disjunctive normal form and $n$ is even.

\item Deciding whether a given $(n,\forall)$-QBF $\Phi$ is valid is $\PiP{n}$-complete.
The problem remains $\PiP{n}$-hard even if the matrix of $\Phi$ is in disjunctive normal form and $n$ is odd,
or in conjunctive normal form and $n$ is even.
\end{enumerate}
\end{proposition}
 
As special cases of these results,  we have  that the satisfiability problem of classical propositional logic
is $\NP$-complete, 
and that 
the validity problem of classical propositional logic is $\CONP$-complete.
Indeed, for a propositional formula $\varphi$ with $V=\var{\varphi}$, $\varphi$ is satisfiable in classical propositional logic iff $\exists V\varphi$ is valid in  quantified propositional logic, and $\varphi$ is valid in classical propositional logic iff $\forall V\varphi$ is valid in quantified propositional logic.

In view of Proposition~\ref{prop:qsat}, we can estimate upper complexity 
bounds for our considered decision problems  
simply by inspecting the quantifier order of the respective encodings.
This can be argued as follows.
First of all, by applying the transformation rules described in 
Proposition~\ref{prop:quantifier-shift}, each of our encodings
can be 
transformed in polynomial time into a closed QBF in prenex form.
Then, by invoking Proposition~\ref{prop:qsat} and observing 
that completeness of a decision problem $D$ for a complexity 
class $C$ implies membership of $D$ in $C$, 
the quantifier order of the resultant QBFs determines in 
which class of the polynomial hierarchy the corresponding 
decision problem lies.

Before dealing with the decision problems associated with equilibrium logic and nested logic programs, we analyse the complexity of the logic of here-and-there.

\begin{theorem}\label{thm:hardness}
\begin{enumerate}

\item\label{thm:hardness:1} Deciding whether a given propositional formula is HT-satisfiable is $\NP$-com\-plete.

\item\label{thm:hardness:2} Deciding whether a given propositional formula is HT-valid is $\CONP$-com\-plete.

\end{enumerate}
\end{theorem}

\begin{proof}
Membership for each of the two tasks in the respective complexity classes follows from the polynomial-time constructible reduction $\PhiHT{\cdot}$ into classical propositional logic. 
Concerning $\NP$-hardness of the HT-satisfiability problem, we show that the $\NP$-hard problem of checking 
whether a given formula in conjunctive normal form is satisfiable in classical propositional logic can be reduced to it in polynomial time.
This can be done as follows.

Let $\varphi$ be a formula in conjunctive normal form, let  $V=\{v_1\commadots v_n\}$ be the set of atoms occurring in $\varphi$, and let $W=\{w_1\commadots w_n\}$ be a set of new atoms.
Furthermore, let ${\varphi}^+$ result from $\varphi$ by replacing
each negative literal 
$\neg v_i$ in $\varphi$ by~$w_i$, and define 
\[
\Trans{\varphi} = \bigwedge_{i=1}^n \big ((v_i \OR w_i) \AND (\neg v_i \OR \neg w_i)\big) \AND {\varphi}^+.
\]

Obviously, $\Trans{\varphi}$ is constructible from $\varphi$ in polynomial time (actually, in linear time).
Furthermore, it holds  that $\varphi$ is satisfiable in classical propositional logic 
iff 
$\Trans{\varphi}$ is HT-satisfiable.

To see this, assume first that $I\subseteq V$ is a model of $\varphi$.
Define 
$$
J_I=I\cup \{ w_i \mid v_i \in V\setminus I\}.
$$
Then, since $\varphi^+$ is an expression without negations, by Part~\ref{lemma:HT:3} of Proposition~\ref{lemma:HT}, we get that $\tuple{J_I,J_I}$ is an HT-model of $\varphi^+$.
But $\tuple{J_I,J_I}$ is also an HT-model of $ \bigwedge_{i=1}^n \big ((v_i \OR w_i) \AND (\neg v_i \OR \neg w_i)\big)$, so it follows that $\Trans{\varphi}$ is HT-satisfiable.

Conversely, consider $I,J\subseteq V\cup W$ such that $\F=\tuple{I,J}$ is an HT-model of 
$\Trans{\varphi}$.
So, $\valF{H}{\Trans{\varphi}}=1$, and therefore $\valF{H}{\bigwedge_{i=1}^n ((v_i \OR w_i) \AND (\neg v_i \OR \neg w_i))}=1$ and
$\valF{H}{\varphi^+}=1$.
Clearly, the former condition entails that $v_i\in I$ iff $w_i\not\in I$, for $1\leq i\leq n$, and by again invoking Part~\ref{lemma:HT:3} of Proposition~\ref{lemma:HT}, from $\valF{H}{\varphi^+}=1$ we get that $\val{I}{\varphi}=1$.
Hence, $I$ is a model of $\varphi$.

It remains to show that checking HT-validity is \CONP-hard.
For this, the $\CONP$-hard problem of deciding whether a given formula in disjunctive normal form is valid in classical propositional logic 
 can be reduced in polynomial time to the problem of deciding HT-validity.

Let $\varphi$ and $\varphi^+$ be as above, and define
\[
\Transv{\varphi} = \big(\bigwedge_{i=1}^n \big ((v_i \OR w_i) \AND (\neg v_i \OR \neg w_i)\big)\Big) \IMPL {\varphi}^+.
\]

Then, analogously to Part~\ref{thm:hardness:1}, it can be shown that $\varphi$ is valid in classical propositional logic iff $\Transv{\varphi}$ is HT-valid.
Moreover, $\Transv{\varphi}$ is constructible from $\varphi$ in linear time.
\end{proof} 

We remark that, since the logic of here-and-there is equivalent to 
the three-valued logic of \citeN{Heyting30} and \citeN{Goedel32}, the above results are also implicit as part of the general characterisation of the complexity of many-valued logics due to \citeN{Mundici87}.

We now turn to the complexity of the main reasoning tasks associated with nested logic programs.

\begin{theorem}\label{thm:complexity:lp}
Both the consistency problem and brave reasoning for nested logic programs are $\SigmaP{2}$-complete, and skeptical reasoning for nested logic programs is $\PiP{2}$-complete.
\end{theorem}

\begin{proof}
The membership conditions for each of the three decision problems are obtained by virtue of Proposition~\ref{prop:qsat} and the encodings given in Corollary~\ref{cor:queries}.
In detail, both the encoding for the consistency problem as well as the encoding for brave reasoning can be transformed in polynomial time into a $(2,\exists)$-QBF.
Hence, these tasks are in $\SigmaP{2}$.
Analogously, the encoding for skeptical reasoning can be transformed in polynomial time into a $(2,\forall)$-QBF, and so skeptical reasoning lies in $\PiP{2}$.

The matching lower bounds are a direct consequence of the complexity of the corresponding reasoning tasks for disjunctive logic programs.
To wit, as shown by \citeN{Eiter95}, both the consistency problem and brave reasoning for disjunctive logic programs are $\SigmaP{2}$-complete, and skeptical reasoning for disjunctive logic programs is $\PiP{2}$-complete.
\end{proof}

For equilibrium logic, analogous complexity bounds are obtained:

\begin{theorem}
Both the consistency problem and brave reasoning for equilibrium logic are $\SigmaP{2}$-complete, and skeptical reasoning for equilibrium logic is $\PiP{2}$-complete.
\end{theorem}

\begin{proof}
The membership conditions are argued in the same way as in the case of logic programs, by using the encodings of Corollary~\ref{cor:eqcons}.

As for the hardness parts, these follow from the respective hardness results from Theorem~\ref{thm:complexity:lp}, by observing that Proposition~\ref{prop:eqmsm} implies that, for any logic program $\Pi$, it holds that $\Pi$ has a stable model iff $\hat{\Pi}$ has an equilibrium model, and $\Pi\nmseq{\varepsilon}\varphi$ iff $\hat{\Pi}\nmseq{\varepsilon}\varphi$, for $\varepsilon\in\{b,s\}$ and any formula $\varphi$.
\end{proof}

We note that $\SigmaP{2}$-hardness of the consistency problem for equilibrium logic was also previously shown by \citeN{Pearce00}.

We now turn to analysing the complexity of checking equivalence between theories and between logic programs.
Again, we start with the case of logic programs.

\begin{theorem}\label{thm:complexity:equivalence:lp}
\begin{enumerate}
\item\label{thm:complexity:equivalence:lp:1} Deciding ordinary equivalence between two given programs is  $\PiP{2}$-complete, and likewise for deciding uniform equivalence.

\item Deciding strong equivalence between two given programs is $\CONP$-complete.

\end{enumerate}

\end{theorem}

\begin{proof}
Again, the membership conditions for each of the three tasks follow from their respective encodings from Corollary~\ref{cor:eqall}.
That is, both the encoding for checking ordinary equivalence as well as the encoding for checking uniform equivalence can be transformed in polynomial time into a $(2,\forall)$-QBF, and the test for strong equivalence is encoded in terms of checking the validity of a formula of classical propositional logic (or, equivalently, checking strong equivalence can be transformed in polynomial time into a $(1,\forall)$-QBF).

$\PiP{2}$-hardness for checking ordinary equivalence between logic programs holds because of the well-known fact that this is already the case for checking equivalence between disjunctive logic programs.\footnote{%
This result follows easily by inspecting the $\SigmaP{2}$-hardness proof 
for the consistency problem for disjunctive logic programs given by \citeN{Eiter95}.
An explicit proof is provided, \egc by \citeN{Oikarinen04}.}
Likewise, checking uniform equivalence between disjunctive logic programs is known to be $\PiP{2}$-complete~\cite{Eiter03}, so $\PiP{2}$-hardness for the corresponding problem for nested problems follows \emph{a fortiori}.

It remains to show that checking strong 
equivalence is $\CONP$-hard. This can be seen as follows:
Consider a formula $\varphi$ and the reduction $\Trans{\varphi}$ from the proof of Theorem~\ref{thm:hardness}. 
Then, one can show that $\varphi$ is unsatisfiable iff the logic program $\{\Trans{\varphi}\LPif\top\}$ is strongly equivalent to the  program $\{\bot\LPif\top\}$. 
Since the problem of checking whether a given formula is unsatisfiable in classical propositional logic is $\CONP$-complete, the $\CONP$-hardness of checking strong equivalence between programs follows.
\end{proof}

For disjunctive logic programs,
$\CONP$-completeness for testing strong equivalence was independently derived by~\citeN{Lin02}, 
and $\CONP$-membership in the case of nested logic programs is also shown by~\citeN{Turner03}.
Also, \citeN{Eiter03} mention that their results for disjunctive logic programs hold for nested programs as well, thus covering the uniform equivalence case in Part~\ref{thm:complexity:equivalence:lp:1} of the above theorem, but no formal account for this claim is provided.

Finally, as a strengthening of Theorem~\ref{thm:complexity:equivalence:lp}, we have the following results for checking equivalence in equilibrium logic.

\begin{theorem}\label{thm:complexity:equivalence:eql}
\begin{enumerate}
\item Deciding whether two given theories of equilibrium logic are ordinarily equivalent is $\PiP{2}$-complete, and likewise for deciding whether they are uniformly equivalent.

\item Deciding whether two given theories of equilibrium logic are strongly equivalent is $\CONP$-complete.

\end{enumerate}

\end{theorem}

The complexity results of this section tell us that our translations adequately match, in a certain sense, the inherent complexity of the respective problems.
In more formal terms, following \citeN{Besnard05}, let us call 
a translation 
${\cal T}_D(\cdot)$, mapping instances of a decision problem $D$ into QBFs,
\emph{adequate} if the following criteria are met:

\begin{enumerate}
\item
For each instance $I$ of $D$,
${\cal T}_D(I)$ is valid iff $I$ is a yes-instance of $D$ (\iec
${\cal T}_D(\cdot)$
is \emph{faithful});
\item
for each instance $I$ of $D$, ${\cal T}_D(I)$ is computable in polynomial time;
and
\item
determining the truth values of the QBFs resulting from ${\cal T}_D(\cdot)$
is not computationally harder than the problem $D$ itself.
\end{enumerate}

It is a straightforward matter to check that all our encodings are indeed adequate in the above sense. More specifically, the following result holds:

\begin{theorem}
All of the encodings described in Corollaries~\ref{cor:eqcons} and \ref{cor:queries}, Theorem~\ref{thm:eqall}, Corollary~\ref{cor:eqall}, and 
Theorem~\ref{cor:UEalt} are adequate.
\end{theorem}

\section{Adding Strong Negation}\label{sec:strong}

Strong negation was introduced into logic by \citeN{Nelson49} as a syntactic counterpart to the
idea of \emph{constructible falsity}. Later, \citeN{Vorobev51} showed how to axiomatise strong negation and provided a reduction technique by which strong negations could be eliminated in favour of additional predicates---a technique later used by \citeN{Gurevich77} to obtain completeness theorems. More recently, strong negation was introduced into logic programming by \citeN{PearceWagner89} and \citeN{Gelfond91}, though the latter called their operator {\em classical negation}. In their answer-set semantics for extended logic programs, \citeANP{Gelfond91} applied the same technique as \citeANP{Vorobev51} to eliminate strong negation and reduce properties of answer sets for extended programs to those of stable models for programs with only one negation.

Equilibrium logic is also defined for theories with two kinds of negation. As before, the basis is the nonclassical logic of here-and-there now augmented by adding a new negation operator, $\negf$, together with the Vorob'ev axioms, see \citeN{Pearce97} and \citeN{Pearce00}.
 This yields a logic called \emph{here-and-there with strong negation}, denoted by ${N_5}$. Interpretations are defined to be here-and-there models as before except that now in each world both atoms and strongly negated atoms may be verified. An atom possibly prefixed by one occurrence of strong negation is called an \emph{objective literal} and a set of objective literals is called \emph{consistent} if it does not contain both $p$ and $\negf{p}$ for some atom $p$.
Hence, an \emph{${N_5}$-interpretation}, $\F$, is an ordered pair $\Iht$ of  consistent sets of objective literals such that as before $I_H \subseteq I_T$. Then, the truth value, $\valF{w}{\varphi}$, of a formula $\varphi$ in a world $w\in\{H,T\}$ in an ${N_5}$-interpretation $\F=\Iht$ is inductively defined as previously, but now with the addition of new clauses for evaluating strong negation.
Formally, for an ${N_5}$-interpretation $\F=\Iht$, a world $w\in\{H,T\}$, and a formula $\varphi$, the {truth value}, $\valF{w}{\varphi}$, {of $\varphi$ in $w$ under $\F$} is given as follows:
\begin{enumerate}
\item if $\varphi=\top$, then $\valF{w}{\varphi} = 1$;

\item if $\varphi=\bot$, then $\valF{w}{\varphi} = 0$;

\item if $\varphi=l$, for some objective literal $l$, then $\valF{w}{\varphi} = 1$ if
      $l\in I_w$, and $\valF{w}{\varphi} = 0$ otherwise;

\item if $\varphi=\neg \psi$, then
      $\valF{w}{\varphi} = 1$ if, for every world $u$ such that $w\leq u$, 
	$\valF{u}{\psi} = 0$, 
      and $\valF{w}{\varphi} = 0$ otherwise;

\item if $\varphi=(\varphi_1\AND \varphi_2)$, then
      $\valF{w}{\varphi} = \MIN(\{\valF{w}{\varphi_1},\valF{w}{\varphi_2}\})$;

\item if $\varphi=(\varphi_1\OR \varphi_2)$, then
      $\valF{w}{\varphi} = \MAX(\{\valF{w}{\varphi_1},\valF{w}{\varphi_2}\})$;

\item if $\varphi=(\varphi_1\IMPL \varphi_2)$, then
      $\valF{w}{\varphi} = 1$ if, for every world $u$ such that $w\leq u$, 
	$\valF{u}{\varphi_1}\leq\valF{u}{\varphi_2}$,  
      and $\valF{w}{\varphi} = 0$ otherwise.

\item if $\varphi=\negf\neg \psi$ or $\varphi=\negf\negf \psi$, then
      $\valF{w}{\varphi} = \valF{w}{\psi}$;

\item if $\varphi=\negf(\varphi_1\AND \varphi_2)$, then
      $\valF{w}{\varphi} = \MAX(\{\valF{w}{\negf\varphi_1},\valF{w}{\negf\varphi_2}\})$;

\item if $\varphi=\negf(\varphi_1\OR \varphi_2)$, then
      $\valF{w}{\varphi} = \MIN(\{\valF{w}{\negf\varphi_1},\valF{w}{\negf\varphi_2}\})$; and

\item if $\varphi=\negf(\varphi_1\IMPL \varphi_2)$, then
      $\valF{w}{\varphi} = \MIN(\{\valF{w}{\varphi_1},\valF{w}{\negf\varphi_2}\})$.
\end{enumerate}
 
An $N_5$-model is defined as in $\HT$, from which we can define an equilibrium model as before \emph{mutatis mutandis}.
To wit, an equilibrium model of a set $T$ of formulas is an ${N_5}$-model $\langle I,I \rangle$ of $T$ such that 
for every proper subset $J$ of $I$, $\langle J,I \rangle$ is not an ${N_5}$-model of $T$.

Enriched in this fashion, equilibrium logic continues to be a conservative extension of the stable-model or answer-set semantics for logic programs. If $\Pi$ is a (nested) logic program possibly containing strong negation, as before we denote by $\hat{\Pi}$ the set of formulas corresponding to the rules of $\Pi$. Then, we have the following relation, extending Proposition~\ref{prop:eqmsm} and shown by \citeN{Lifschitz01}:

\begin{proposition}
\label{prop:eqas}
For any program $\Pi$, $I$ is an answer set of $\Pi$
iff 
$\langle I, I \rangle$ is an equilibrium model of $\hat{\Pi}$.
\end{proposition} 

Our previous transformations can easily be extended to cover strong negation by deploying the reduction technique of Vorob'ev and Gurevich. The method involves two steps and is carried out as follows. First, let us say that a formula of ${N_5}$ is in \emph{reduced form} if any occurrence of strong negation appears directly in front of an atom. The first step is then, given any formula $\varphi$, to convert $\varphi$ into an equivalent formula $\varphi^*$ in reduced form. The translation `$\cdot^*$' is inductively defined as follows:

\begin{enumerate}
\item
$ p^* = p$ and $(\negf p)^* = \negf p$, if $p$ is an atom;

\item
$(\neg \varphi)^* = \neg \varphi^*$;

\item
$(\varphi \circ \psi)^* = \varphi^* \circ \psi^*$, for $\circ\in\{\AND,\OR,\IMPL\}$;

\item
$(\negf \neg \varphi)^* = (\negf \negf \varphi)^* = \varphi^*$;

\item
$\negf (\varphi \AND \psi)^* = \negf \varphi^* \OR \negf \psi^*$;

\item
$\negf (\varphi \OR \psi)^* = \negf \varphi^* \AND \negf \psi^*$; and

\item
$\negf (\varphi \IMPL \psi)^* = \varphi^* \AND \negf \psi^*$.
\end{enumerate}
It is easy to establish that, for any formula $\varphi$, $\varphi^*$ is in reduced form and, moreover, $\varphi \IFF \varphi^*$ is provable in ${N_5}$.
Furthermore, it can be shown that computing $\varphi^*$ from $\varphi$ is polynomial in $\lc{\varphi}$. 

In the second step, we extend the language by adding, for each atom $p$, a new atom $\clneg{p}$ not in the language. 
Let $\varphi$ be a formula.
Then, set $c_\varphi =
\bigwedge_{p\in \var{\varphi}} \neg (p \AND\clneg{p})$. Furthermore, for any ${N_5}$-interpretation $\F=\Iht$, let $\overline{\F}$ be the HT-interpretation obtained from $\F$ by replacing each occurrence of a strongly negated literal $\negf p$ by $\clneg{p}$. Finally, for any formula $\varphi$ of ${N_5}$, let $r(\varphi)$ be the formula that results from replacing each occurrence of $\negf p$ in $\varphi^*$ by $\clneg{p}$. Clearly, $r(\varphi)$ is a formula of $\HT$.

Then, we have:

\begin{lemma}
For any formula $\varphi$ of ${N_5}$ with $V=\var{\varphi}$, any $N_5$-interpretation $\F=\Iht$, and any world $w\in\{H,T\}$, 
$\varphi$ is true in $w$ under $\F$ iff $c_\varphi\AND r(\varphi)$ is true in $w$ under $\overline{\F}$.
\end{lemma}

\begin{proof}
By the validity of $\varphi \IFF \varphi^*$ in $N_5$, we need only show that 
$\valF{w}{\varphi^*}=1$ iff $\valFs{w}{c_\varphi\AND r(\varphi)}=1$.
This follows by an easy induction on $\varphi^*$. The cases for objective literals follow from the definition of $\overline{\F}$. 
For the other connectives, note that, since $\varphi^*$ is in reduced form, we only need to check the satisfaction conditions for $\AND$, $\OR$, $\IMPL$, and $\neg$. But these are the same for $\F$ as for $\overline{\F}$. Lastly, the fact that $\F$ is a (consistent) interpretation guarantees that $c_\varphi$ is true in $w$ under $\overline{\F}$.
\end{proof}

Let $\F$ be an HT-interpretation whose atoms belong to $V \cup \overline{V}$, where $\overline{V} = \{ \clneg{p} \mid p \in V \}$. 
We define $\F^\dagger$ to be the ${N_5}$-interpretation obtained by uniformly replacing each occurrence of $\clneg{p}$ for an atom $p$ by $\negf p$. By the definition of equilibrium model and the above lemma, we obtain:

\begin{theorem}
For any formula $\varphi$ of ${N_5}$, the following conditions hold: 

\begin{enumerate}
\item If an ${N_5}$-interpretation $\F$ is an equilibrium model of $\varphi$, then $\overline{\F}$ is an equilibrium model of $c_\varphi \AND r(\varphi)$. 

\item If an HT-interpretation $\F$ is an equilibrium model of $c_\varphi \AND r(\varphi)$, then $\F^\dagger$ is an equilibrium model of $\varphi$.

\end{enumerate}
\end{theorem}

In this manner, we can extend the previous reduction technique for equilibrium logic to the enriched system with strong negation. We need only convert each formula $\varphi$ to reduced form and add the conjunction $c_\varphi=\bigwedge_{p\in \var{\varphi}} \neg (p \AND\clneg{p})$. 
The same applies to logic programs with strong negation. 
However, in the latter case, under the answer-set semantics of \citeN{Lifschitz99}, it is assumed that strong negations are placed directly before atoms, hence the corresponding formulas are already in reduced form. We can therefore dispense with step one. The procedures $\PhiE{\hat{\Pi}}$ and the optimised variant $\PhiEn{\Pi}$, for each program $\Pi$, are extended to the notion of answer sets again by adding the conjunction $\bigwedge_{p\in \var{\Pi}} \neg (p \AND\clneg{p})$. 

An additional frequently used notion is that of a \emph{complete} answer set.
An answer set $I$ of a logic program $\Pi$ is complete iff, 
for each atom $p$ in $\Pi$, either $p$ or $\negf p$ is contained in $I$. 
To deal with complete answer sets, it suffices to add a second conjunction,
$\bigwedge_{p\in \var{\Pi}} (p \OR \clneg p)$.
Extensions to encode brave and skeptical reasoning 
for logic programs containing strong negation, as well as 
encoding consequence relations defined in terms of complete answer sets, 
can be defined analogously.

\section{Related Work}\label{sec:rel}

Capturing one formalism in terms of another is a natural issue for theoretical study, and accordingly various translatability results have appeared in the logic-programming literature (\egc let us recall the various embeddings of logic programs into autoepistemic logic, which we mentioned in Section~\ref{sec:circ}).

In what follows, we first discuss studies dealing with translations to classical logic---these are clearly closest in spirit to our work.
Afterwards, we consider translations from one syntactic subclass of programs to another. Finally, we give some pointers concerning work on equivalence checking.

\subsection{Translations of Logic Programs into Classical Logic}

In face of the inherent complexity of equilibrium logic and nested logic programs, and assuming the usual proviso that the polynomial hierarchy does not collapse, translations of reasoning tasks associated with any of these languages into classical propositional logic usually belong to one of the following two categories: 
\begin{itemize}
\item either one is interested in \emph{efficient} translations, in which case these can be realised only for subclasses of programs or theories matching the complexity of classical logic, 
\item or else one deals with translations for \emph{arbitrary} programs or theories, for which an exponential blow-up of the translations must be faced in the worst case.
\end{itemize}
Concerning the first kind of results, early efforts in that direction involve the work of \citeN{Clark78}, defining the \emph{completion} of a program, and of \citeN{Fages94}.
\citeN{Ben-Eliyahu94} defined a syntactic subclass of disjunctive logic programs, viz.\ \emph{head-cycle free programs}, along with a translation of these programs into theories of classical logic constructible in polynomial time.
This translation was optimised by \citeN{Janhunen04} for {normal logic programs},
and \citeN{Linke04} provide a generalisation of the notion of head-cycle freeness to nested logic programs, together with an extension of the translation of \citeN{Ben-Eliyahu94}.
As well, \citeN{Linke04} give a generalisation of Fages' theorem to 
so-called \emph{tight nested logic programs}, encompassing earlier results due to Erdem and Lifschitz~\shortcite{Lifschitz01a,Erdem03a}. 

A prominent technique for realising translations of programs without any additional syntactic proviso into classical logic, but instead accepting an exponential blow-up in the worst case, are based on adding to Clark's completion so-called \emph{loop formulas}, guaranteeing equivalence
between the stable models of the given program and the classical models of the
resultant theory.
This idea was first pursued by~\citeN{Lin02a}
for normal programs and later
extended by~\citeN{Lee:Lifschitz:03}
for disjunctive programs with nested formulas in rule bodies.
Further extensions of this approach were put forth 
by \citeN{Lee05} and \citeN{Gebser06}.
Implementations of the loop-formula approach 
are the {\tt ASSat} system~\cite{Lin02a} and 
the {\tt CMODELS} system~\cite{Lierler05}.
An interesting relation between the present QBF encodings and loop formulas has been analysed by \citeN{Ferraris06}.

\subsection{Translations between Different Program Classes}

Given that the main reasoning tasks associated with equilibrium logic and nested logic programs have the same complexity as the corresponding problems for disjunctive logic programs entails that the former can be efficiently reduced to the latter.
A polynomial translation from nested logic programs to disjunctive ones has been realised by \citeN{Pearce02}, based on a labelling technique and exploiting a translation from \emph{generalised disjunctive logic programs} to disjunctive programs due to \citeN{Janhunen01}.\footnote{In a generalised disjunctive logic program, rule heads are disjunctions of literals and rule bodies are conjunctions of literals.} Although already \citeN{Lifschitz99} provide a translation from nested programs to disjunctive ones, their translation is exponential in the worst case.
Furthermore, the translation by \citeN{Pearce02} satisfies \emph{strong faithfulness}, expressing that, for
every program $\Pi_1$ and $\Pi_2$, there is a one-to-one
correspondence between the answer sets of $\Pi_1\cup\Pi_2$ and sets of form
$I\cap\var{\Pi_1\cup\Pi_2}$, where $I$ is an answer set of $\rho(\Pi_1)\cup\Pi_2$, with $\rho(\Pi_1)$ being the translation of $\Pi_1$.
Note that strong faithfulness is a refined version of strong equivalence, taking differing alphabets into account, as the translation $\rho$ introduces new atoms.

As for translating equilibrium logic to disjunctive logic programs, this was put forth 
by \citeN{Cabalar05}, making use of a technique due to \citeN{Osorio05}.
In subsequent work, \citeN{Cabalar06} show that each theory can also 
be rewritten into a \emph{strongly} equivalent nested logic program 
(albeit with an exponential blow-up in general)  and \citeN{cabalar07} discuss the issue of obtaining a \emph{minimal} such program, involving steps similar to the well-known Quine-McCluskey algorithm for minimising propositional formulas~\cite{quine52,McCluskey56}. 

Attention was also given in the literature to translations of disjunctive logic programs to normal programs.
To wit,
\citeN{Ben-Eliyahu94} provide a method to transform head-cycle free disjunctive programs
equivalently into normal programs by shifting head atoms into 
the body.\footnote{This shifting method is also discussed by \citeN{GelLifPrzTru91} and \citeN{Dix96}.}
For instance, a rule of form $p\OR q\LPif r$ is replaced by this method 
by the two rules $p\LPif r\AND \neg q$ and $q\LPif r\AND \neg p$.
This shifting technique was later extended by \citeN{Linke04} to head-cycle free nested logic programs, thus providing a translation of the latter to normal programs.
It is worth mentioning that this technique
partly relies on a translation due to \citeN{yoyuzh03} and extends ideas by~\citeN{Inoue98}.
A general investigation under what conditions disjunctions can be eliminated from disjunctive logic programs was carried out by \citeN{Eiter04}.
This was done for ordinary, uniform, and strong equivalence, and makes use of the above shifting technique.

\subsection{Program Equivalence}

We now turn to work regarding equivalence checking. 
Besides the notions analysed in this paper, more refined ones have been introduced in the literature as well.
On the one hand, \citeN{Woltran04} defines \emph{relativised} notions of uniform and strong equivalence for disjunctive logic programs, in which the alphabet of the rules added for the comparison is taken into account.
On the other hand, this investigation was subsequently extended by Eiter, Tompits, and Woltran \citeyear{Eiter05}, defining a general framework for specifying program correspondence along with model-theoretic characterisations and complexity results.
This framework allows for the specification of parameterised equivalence notions, taking, \egc projected answer sets into account where auxiliary atoms are ignored for program comparison.
We note that the latter feature is important towards realising modular programming, which is in turn addressed by \citeN{Janhunen06} in connection with program comparison under \emph{modular equivalence}, a form of ordinary equivalence taking local atoms into account.
Concerning the complexity of relativised uniform and strong equivalence, this was thoroughly analysed by \citeN{Eiter04TR}.

Related to our encodings for uniform equivalence is the work by Oetsch, Tompits, and Woltran~\shortcite{Oetsch07b}, dealing with relativised uniform equivalence under projection between disjunctive logic programs under the stable-model semantics, a generalisation of uniform equivalence within the framework of \citeN{Eiter05}.
They discuss a model-theoretic characterisation for this kind of equivalence, which is in some sense orthogonal to the model theory for uniform equivalence based on UE-models~\cite{Eiter03}, and provide encodings for equivalence testing by means of quantified propositional logic.
Likewise complementing our results from Section~\ref{sec:eq} is the investigation by \citeN{gebser-etal-foiks08}, who introduce alternative characterisations for strong and uniform equivalence in terms of \emph{unfounded sets}, the key constituents underlying the well-founded semantics~\cite{gerosc91a}, and, based on these results, discuss encodings of testing strong and uniform equivalence between disjunctive logic programs in terms of standard and quantified propositional logic, respectively.

Another generalisation of strong, uniform, and ordinary equivalence, proposed by \citeN{Woltran07}, is \emph{head-body relativised equivalence} (subsequently termed \emph{hyperequivalence}~\cite{tw-aaai08}), 
which has as additional parameter the possibility to specify different alphabets for heads and bodies of rules.
Extending the investigations of \citeN{Eiter04}, conditions under which disjunction and negation can be eliminated under hyperequivalence are studied by P\"uhrer, Tompits, and Woltran~\shortcite{puehrer-iclp08}.

\citeN{inou-saka-04} define equivalence under program \emph{updates}, and 
generalised characterisations of strong and uniform equivalence
for language extensions like 
cardinality constraints
and preferences were put forth by \citeN{Liu05} and by \citeN{Faber05}, respectively. 
Studies of strong equivalence for other extensions of logic programs include, \egc the work of  \citeN{faber-etal-kr08} for programs with \emph{ordered disjunction}~\cite{brew-2002-aaai,brew-etal-2004-ci} and of \citeN{cabalar-jelia08} for \emph{temporal equilibrium logic}~\cite{cabalar-eurocast07}.
Finally, the notion of \emph{synonymous theories} in the context of equilibrium logic was introduced by \citeN{Pearce04c}.

Concerning work on implementational aspects, let us first recall that our axiomatisations straightforwardly enable the development of provers for equilibrium logic using extant QBF solvers as back-end inference engines.
In any case, a concrete implementation for equilibrium logic was developed by 
\citeN{Valverde04}, relying on a tableau calculus due to \citeN{Pearce00a}.
As well, an implementation for nested logic programs, based on the reduction to disjunctive logic programs put forth by \citeN{Pearce02} and using the underlying solver \texttt{DLV}~\cite{leon-etal-2002-dlv}, is discussed by \citeN{sasctowo03a}.
A reduction approach to disjunctive logic programs is also realised by several systems for equivalence checking.
To begin with, the system \texttt{dlpeq} implements a tester for ordinary and strong equivalence. 
It was first developed for normal logic programs \cite{Janhunen02} and afterwards extended to disjunctive programs \cite{Oikarinen04}.
A similar idea to decide strong equivalence using the language of logic programs itself was independently
suggested by \citeN{Turner03} and some aspects are carried out 
by \citeN{Eiter03a} as well.
Another implementational approach is pursued by the system \texttt{SELP} \cite{Lin05} for checking strong equivalence, using the reductions to classical propositional logic described by \citeN{Lin02}  (see also our discussion in Section~\ref{sec:eq} about these translations), and also by the work of Eiter, Faber, and Traxler \citeyear{EFT05}.
Finally, the family of equivalence notions from the framework of 
Eiter, Tompits, and Woltran \citeyear{Eiter05} is implemented by the system \texttt{cc}$\top$ \cite{Oetsch06b,Oetsch06c,Oetsch07c}, using reductions to quantified propositional logic put forth by \citeN{Tompits05} and \citeN{Oetsch07b}, similar to the reduction approach followed in the present paper.
We note that testing strong and uniform equivalence is one of the suite of notions handled by \texttt{cc}$\top$, and the reduction used for computing strong equivalence in this system is equivalent to the one used by \texttt{SELP}.

\section{Concluding Remarks}

In this paper, we discussed how different reasoning problems 
in the context of equilibrium logic and nested logic programs  
can be expressed in a uniform framework by means of quantified
propositional logic.
We have started by introducing basic formulas that are used as building
blocks for modeling differing reasoning tasks.
Our results thus provide new axiomatisations for the formalisms under consideration and contribute to the analysis of the logical foundations of answer-set programming and its extensions.

The overall approach has several benefits.
First of all, it allows us to compare different problems in a single formal language.
This is of particular interest for the
three different notions of equivalence considered here, viz.\ ordinary, uniform, and strong equivalence.
Moreover, the axiomatisations imply upper complexity bounds
in a direct manner for the problems under consideration, which we all strengthened to completeness results, thus providing strict complexity bounds.
Lastly, the axiomatisations provide executable specifications that can be
fed into extant QBF solvers, yielding prototypical implementations for the encoded problems.
In view of the considerable sophistication offered by current QBF solvers, one obtains a viable approach for rapid prototyping.

Our work is situated within a propositional language, yet liftings to the nonground case are possible.
Indeed, both the logic of here-and-there and equilibrium logic have been extended to the first-order case, introducing \emph{quantified here-and-there} (QHT) and \emph{quantified equilibrium logic}~\cite{pearce-valverde-sl05}, respectively, and our encodings directly lift to these extensions, yielding characterisations of QHT in terms of first-order logic as well as characterisations of quantified equilibrium logic in terms of second-order logic.
These encodings are discussed by \citeN{Ferraris:Lee:Lifschitz:07} and \citeN{pearce-valverde-08} in the context of defining stable models for general first-order theories.
However, as for propositional programs, this newly defined stable-model semantics for first-order theories basically amounts to quantified equilibrium logic, \iec a stable model of a first-order theory in the sense of \citeN{Ferraris:Lee:Lifschitz:07} corresponds to an equilibrium model in quantified equilibrium logic and vice versa.
It is worth noting that this new definition of a stable model does not coincide with the usual one when applied to a nonground program because the unique-names assumption is no longer assumed---indeed, only stable models which are Herbrand models coincide with stable models in the usual sense.

Emanating from these works, there is currently an active line of research to lift well-known results about the traditional stable-model semantics to its generalised version (\iec in effect, to quantified equilibrium logic).
For instance, work in this direction includes 
\begin{itemize}
\item the extension of counting and choice rules to the new definition~\cite{lee-lifschitz-aaai08},

\item a characterisation of strong equivalence between first-order theories in terms of a version of QHT~\cite{lifschitz-lpnmr07}, similar to the correspondence between strong equivalence and HT in the propositional case studied earlier by the same authors \cite{Lifschitz01}, and

\item model-theoretic characterisations of uniform equivalence between theories of quantified equilibrium logic~\cite{fink08}.

\end{itemize}

Concerning more general equivalence relations in the context of nonground programs, we mention the work by \citeN{oetsch-tompits-iclp08}, where the program correspondence framework of Eiter, Tompits, and Woltran~\shortcite{Eiter05} is extended to the nonground case and characterisations in terms of second-order logic for relativised uniform equivalence with projection between nonground disjunctive logic programs are given.
An interesting issue for future work would be to extend these characterisations, as well as characterisations for relativised strong equivalence~\cite{Eiter05,Tompits05}, to full quantified equilibrium logic.

\paragraph{Acknowledgments.}

We would like to thank the anonymous referees for their valuable comments which helped to improve the paper.
This work was supported by the Spanish Ministry of Education and Science (MEC, later MICINN) under projects TIC-2003-9001-C02, TIN2006-15455-CO3, and Agreement Technologies CSD2007-00022, and by the Austrian Science Fund (FWF)
under grant P18019. 

\appendix
\section{Proof of Lemma~\ref{thm:HTSat}}\label{app:proof:HT}
The proof proceeds on the number of connectives $\AND$, $\OR$, and $\IMPL$ in $\varphi$
which are not in the scope of a negation. Denote this number
by $\lcn{\varphi}$.
First, observe that, for $I_H,I_T\subseteq V$, 
the pair $\langle I_H, I_T\rangle$ is an HT-interpretation
(\iec $I_H\subseteq I_T$) 
iff $I_H\cup I_T'$ is a model of $V\leq V'$.

\medskip
\noindent
\emph{Induction Base.} 
Assume $\lcn{\varphi}=0$. Then,  $\varphi$ is either an atom, one of $\top$ or $\bot$, or of form $\neg \psi$, for some formula $\psi$.
If $\varphi$ is one of $\top$ or $\bot$, then the statement holds trivially.

Consider now the case that $\varphi=p$, for some atom $p$.
Then, $V=\{p\}$ and $\PhiT{p}=p$. Hence,
$
\PhiHT{\varphi} = (p \IMPL p') \AND p$.

Assume that $\F=\langle I_H, I_T\rangle$ is an HT-model
of $\varphi$, \iec we have that $I_H\subseteq I_T\subseteq V$ and $p\in I_H$. 
From this, we immediately get that $I_H=I_T=V=\{p\}$,
and so $I_H\cup I_T'=\{p,p'\}$, which is clearly a model of $\PhiHT{\varphi}$. Conversely, if $I_H\cup I_T'\subseteq V\cup V'$ is a model 
of $\PhiHT{\varphi}=(p \IMPL p') \AND p$, 
then clearly $I_H=I_T=\{p\}$.
Hence, $\F=\Iht$ is an HT-interpretation.
Moreover, in view of $p\in I_H$, $\F$ is an HT-model of $\varphi$.

Now assume that $\varphi=\neg \psi$, for some formula $\psi$.
By definition of \PhiT{\cdot},
$
\PhiHT{\varphi}=(V\leq V') \AND \neg \psi'$.

Suppose that $\F=\langle I_H, I_T\rangle$ is an HT-model of
$\varphi=\neg \psi$.
Then,  $\valF{w}{\psi} = 0$, for each $w\in\{H,T\}$.
By Proposition~\ref{lemma:HT}, $\valF{T}{\psi}=0$ implies $\val{I_T}{\psi}=0$.
Hence, $\psi'$ is false under $I_T'$,
and consequently $\neg\psi'$ is true under $I_H\cup I'_T$.
Moreover, since $\F$ is an HT-interpretation, $V\leq V'$ is true under $I_H\cup I'_T$.
Thus, $I_H\cup I'_T$ is a model of $\PhiHT{\varphi}$.

Conversely, if $I_H\cup I'_T$ is a model of $\PhiHT{\varphi}=(V\leq V')\AND \neg\psi'$, then
$\psi'$ is false under $I_T'$.
From this, a simple renaming yields $\val{I_T}{\psi} = 0$.
Now, since  $I_H\cup I'_T$ is also a model of $(V\leq V')$, $\F=\Iht$ must be an HT-interpretation, and therefore  Part~\ref{lemma:HT:1} of Proposition~\ref{lemma:HT} implies $\valF{T}{\psi} = 0$.
But Part~\ref{lemma:HT:2} of the same proposition yields $\valF{H}{\psi} = 0$.
Consequently, $\valF{H}{\neg\psi} = 1$, and so $\F$ is an HT-model of $\varphi$.

\medskip
\noindent
\emph{Induction Step.}
 Assume $\lcn{\varphi}>0$, and let the statement hold
for all formulas $\psi$ such that  $\lcn{\psi}<\lcn{\varphi}$.
We have to consider several cases, depending on the structure of $\varphi$.

\begin{description}
\item[\rm\emph{Case 1.}] Assume that $\varphi=(\varphi_1\AND\varphi_2)$.
Then,
$\PhiHT{\varphi}  = (V \leq V') \AND \PhiT{\varphi_1\AND\varphi_2}$,
with $\PhiT{\varphi_1\AND\varphi_2}=\PhiT{\varphi_1}\AND\PhiT{\varphi_2}$.
Since $V\leq V'$ is equivalent in classical logic to $(V_1\leq V_1')\AND (V_2\leq V_2')$, where $V_i=\var{\varphi_i}$, for
$i=1,2$, it follows that
$\PhiHT{\varphi}$ is equivalent in classical logic to
$$
\big((V_1 \leq V_1') \AND \PhiT{\varphi_1}\big)\AND\big((V_2 \leq V_2') \AND \PhiT{\varphi_2}\big),
$$
which in turn represents $\PhiHT{\varphi_1}\AND\PhiHT{\varphi_2}$.

Suppose now that $\F=\langle I_H, I_T\rangle$ is an HT-model of $\varphi=(\varphi_1\AND\varphi_2)$. Hence, $\F$ is an HT-model of $\varphi_1$ and $\varphi_2$. Since $\lcn{\varphi_i}<\lcn{\varphi}$, for $i=1,2$, by induction hypothesis we get that $I_H\cup I'_T$ is a model of both $\PhiHT{\varphi_1}$ and $\PhiHT{\varphi_2}$, and thus also of $\PhiHT{\varphi_1}\AND\PhiHT{\varphi_2}$. It follows that $I_H\cup I'_T$ is a model of $\PhiHT{\varphi}$.
The proof of the converse direction proceeds analogously.

\item[\rm\emph{Case 2.}] Assume that $\varphi=(\varphi_1\OR\varphi_2)$. Similar to the above,  $\PhiT{\varphi}=\PhiT{\varphi_1}\OR\PhiT{\varphi_2}$, hence $\PhiHT{\varphi}$ is given by
\begin{equation}\label{eq:1}
(V \leq V') \AND (\PhiT{\varphi_1}\OR \PhiT{\varphi_2}).
\end{equation}
By taking $V_i=\var{\varphi_i}$, for
$i=1,2$, it follows  that (\ref{eq:1}) is classically equivalent to
$$
(V \leq V') \AND \Big(\big((V_1\leq V_1')\AND\PhiT{\varphi_1}\big)\OR \big((V_2\leq V_2')\AND\PhiT{\varphi_2}\big)\Big),
$$
which represents  $(V \leq V')\AND (\PhiHT{\varphi_1}\OR\PhiHT{\varphi_2})$.

Suppose now that $\F=\langle I_H, I_T\rangle$ is an HT-model of $\varphi=(\varphi_1\OR\varphi_2)$. We get that $\F$ is an HT-model of $\varphi_1$ or of $\varphi_2$. Without loss of generality, assume that $\F$ is an HT-model of $\varphi_1$.
Since $\lcn{\varphi_1}<\lcn{\varphi}$, by induction hypothesis it follows that $I_H\cup I'_T$ is a model of $\PhiHT{\varphi_1}$. Hence, $I_H\cup I'_T$ is also a model of $\PhiHT{\varphi_1}\OR\PhiHT{\varphi_2}$.
Furthermore, since $\F$ is an HT-interpretation, $I_H\cup I'_T$ is a model of $V\leq V'$. 
Therefore, $I_H\cup I'_T$ is a model of $(V \leq V')\AND (\PhiHT{\varphi_1}\OR\PhiHT{\varphi_2})$. Since the last formula is equivalent to $\PhiHT{\varphi}$, we obtain that $I_H\cup I'_T$ is a model of $\PhiHT{\varphi}$. The converse direction follows in a similar way.

\item[\rm\emph{Case 3.}] Assume that $\varphi=(\varphi_1\IMPL\varphi_2)$. Then,
$$
\PhiHT{\varphi}= (V\leq V') \AND (\PhiT{\varphi_1}\IMPL\PhiT{\varphi_2})\AND
(\varphi'_1 \IMPL \varphi'_2).
$$
It is easy  to see that $\PhiHT{\varphi}$ is equivalent in classical logic to
\begin{equation}\label{eq:main:impl}
(V\leq V') \AND \Big( \big( (V_1\leq V_1') \AND \neg \PhiT{\varphi_1} \big)
\OR \big((V_2\leq V_2') \AND \PhiT{\varphi_2}\big) \Big) \AND (\varphi'_1 \IMPL \varphi'_2),
\end{equation}
where $V_i=\var{\varphi_i}$, for $i=1,2$. 
 
Assume that $\F=\langle I_H, I_T\rangle$ is an HT-model of $\varphi$. 
This means that
(i)~$\valF{H}{\varphi_1} = 0$ or $\valF{H}{\varphi_2} = 1$, and~(ii) $\valF{T}{\varphi_1} = 0$ or $\valF{T}{\varphi_2} = 1$.
We show that each conjunct of (\ref{eq:main:impl}) is true under $I_H\cup I'_T$.

Clearly, $V\leq V'$ is true under $I_H\cup I'_T$ because $\F$ is an HT-interpretation. Moreover, Part~\ref{lemma:HT:1} of Proposition~\ref{lemma:HT} implies that (ii) is equivalent to the condition that $\val{I_T}{\varphi_1}=0$ or $\val{I_T}{\varphi_2}=1$. From the latter we obtain by a simple renaming that $I_H\cup I'_T$ is a model of $(\varphi'_1 \IMPL \varphi'_2)$. 
It remains to show that $I_H\cup I'_T$ is a model of
\begin{equation}\label{eq:main:impl:sub}
\big( (V_1\leq V_1') \AND \neg \PhiT{\varphi_1} \big) 
\OR \big((V_2\leq V_2') \AND \PhiT{\varphi_2}\big).
\end{equation}

From (i) we know that either $\valF{H}{\varphi_1} = 0$ or $\valF{H}{\varphi_2} = 1$, or both.
Similar to arguments in the induction base, one can show that $\valF{H}{\varphi_1} = 0$ implies that   $I_H\cup I'_T$ is a model of $(V_1\leq V_1') \AND \neg \PhiT{\varphi_1}$.
Now assume that $\valF{H}{\varphi_2} = 1$, \iec $\F$ is an HT-model of $\varphi_2$. Since $\lcn{\varphi_2}<\lcn{\varphi}$, by induction hypothesis we get that $I_H\cup I'_T$ is a model of $(V_2\leq V_2') \AND \PhiT{\varphi_2}$. Hence, in either case, we can conclude that $I_H\cup I'_T$ is a model of~(\ref{eq:main:impl:sub}). This shows that if $\F$ is an HT-model of $\varphi$, then $I_H\cup I'_T$ is a model of $\PhiHT{\varphi}$. The proof of the converse direction follows in a similar fashion. $\mathproofbox$
\end{description}

\bibliography{d}

\end{document}